\documentclass[preprint,12pt,graphics]{elsarticle}

\usepackage{amssymb}

\usepackage{lineno}

\newcommand{\vol}[2]{$\,$\rm #1\rm , #2}

\newcommand\apj{{Astrophysical Journal}}%
\newcommand\apss{{Ap\&SS}}%
\newcommand\aap{{A\&A}}%
\newcommand\mnras{{Monthly Notices of the Royal Astronomical Society}}%
\newcommand\procspie{{Proc.~SPIE}}%

\journal{Astroparticle Physics}
\begin{document}

\begin{frontmatter}

\title{Scientific Prospects for Hard X-ray Polarimetry}


\author[a1]{H. Krawczynski \corref{ca}}
\cortext[ca]{Corresponding author.}
\ead{krawcz@wuphys.wustl.edu, Tel. 314 935 8553, Fax. 314 935 6219}
\author[a1]{A.~Garson III}
\author[a1]{Q.~Guo}
\author[a2]{M.~G.~Baring}
\author[a1]{P.~Ghosh}
\author[a1]{M.~Beilicke}
\author[a1]{K.~Lee}
\address[a1]{Washington University in St. Louis,
   Department of Physics and McDonnell Center for the Space Sciences,
   1 Brookings Dr., CB 1105,
   St Louis, MO 63130}
\address[a2]{
Rice University,
Department of Physics and Astronomy - MS 108,
P.\ O.\ Box 1892,
Houston, Texas 77251}

\begin{abstract}
X-ray polarimetry promises to give qualitatively new information about high-energy sources. 
Examples of interesting source classes are binary black hole systems, rotation and accretion 
powered neutron stars, Microquasars, Active Galactic Nuclei and Gamma-Ray Bursts. 
Furthermore, X-ray polarimetry affords the possibility for testing fundamental physics, e.g.\ 
to observe signatures of light 
bending in the strong gravitational field of a black hole, to detect third order 
Quantum Electrodynamic effects in the magnetosphere of Magnetars, and to perform 
sensitive tests of Lorentz Invariance.  In this paper we discuss scientific drivers of   
hard ($>$10 keV) X-ray polarimetry emphasizing how observations in the hard band 
can complement observations at lower energies (0.1 - 10 keV). Subsequently, we describe 
four different technical realizations of hard X-ray polarimeters suitable for
small to medium sized space borne missions, and study their performance in the 
signal-dominated case based on Monte Carlo simulations. We end with confronting 
the instrument requirements for accomplishing the science goals with the 
capabilities of the four polarimeters.
\end{abstract}
\begin{keyword}
Hard X-ray Polarimetry \sep Instrumentation \sep Compton Effect 
\sep X-ray detectors \sep Binary Black Holes \sep Neutron Stars 
\sep Pulsars \sep Gamma Ray Bursts \sep Active Galactic Nuclei \sep
Blazars \sep Lorentz Invariance Violation
\end{keyword}
\end{frontmatter}
\section{Introduction}
\label{intro}
Compared to observations in other parts of the electromagnetic spectrum, X-ray observations 
are of particular interest for the study of mass accreting black holes and  
neutron stars because these objects are X-ray bright and the X-rays originate 
very close to the compact objects. It is thus not surprising that X-rays are 
key to explore the properties of these objects. Whereas several X-ray imaging, spectroscopy, and timing missions have made 
spectacular discoveries over the last three decades 
\citep{Giac:79,True:83,Weiss:02,Stru:01,Jaho:96,Mi:07}, only one dedicated X-ray polarimetry 
mission has been launched so far. One of the reasons is that it is difficult to measure 
the polarization of an X-ray beam: 
whereas the arrival direction, arrival time, and energy of individual photons 
can be measured with extremely high accuracy, many hundreds of photons are needed to make even rough
measurements of the three Stokes parameters $P$, $Q$, and $V$ which characterize the polarization
properties of an X-ray beam. The only dedicated X-ray polarimetry mission
to date OSO-8 \citep{Novi:75} detected a 2.6~keV and 5.2~keV polarization of the X-rays
from the Crab Nebula of $\sim$20\% and a polarization angle aligned around 
30 degrees oblique to the X-ray jet \citep{Weis:78}. 
For Cyg X-1 weak evidence for polarization on a level of a few percent was found \citep{Silv:79}; 
for other galactic compact objects upper limits on the polarization degree of a few 10 percent were measured \citep{Silv:79,Hugh:84}. 
Recently, two instruments on the {\it INTEGRAL} satellite were used to constrain 
the polarization of the hard X-ray emission from the Crab Nebula. 
Based on the analysis of data from the SPI instrument (SPectrometer on {\it INTEGRAL}), 
\citet{2008Sci...321.1183D} report tentative evidence for a 46\%$\pm$10\% polarization 
degree of the 100 keV-1~MeV emission. The analysis of 200 keV - 1 MeV data from the
IBIS instrument (Imager on Board the {\it INTEGRAL} Satellite) indicates an
even higher polarization fraction \citep{Foro:08}. The polarization direction seems 
to be aligned with the orientation of the X-ray jet at these 
energies \citep{2008Sci...321.1183D,Foro:08}. 
Models predict that galactic sources (e.g.\ binary black holes 
and neutron stars) and extragalactic sources (e.g.\ blazars, Gamma-Ray Bursts, GRBs) exhibit linear 
polarization degrees of a few percent and a few tens of percent, respectively, slightly 
below the sensitivity of the OSO 8 experiment. A mission with an order of magnitude 
improved sensitivity over OSO 8 should thus be able to detect the polarization 
of many objects, and to provide spectacular galactic and extragalactic results.

Recent technological progress namely photo-electron tracking gas detectors \citep{Hill:07,Costa:08} 
have opened up the possibility to design small X-ray missions with more than
two orders of magnitude better polarization sensitivities than OSO 8. The Gravity and 
Extreme Magnetism SMEX ({\it GEMS}) mission \citep{Swan:10} with excellent polarimetry 
sensitivity in the 2-10 keV energy band has recently 
been approved as a NASA mission. {\it GEMS} is projected to achieve a Minimum 
Detectable Polarization (MDP) degree of about 3\% for a mCrab source and an 
integration time of 1000 ksec. {\it GEMS} will have a single-photon energy resolution 
of between 15\% and 20\% Full Width Half Maximum (FWHM) and no imaging capabilities. 
At 0.5 keV {\it GEMS} will fly a student polarimeter with modest sensitivity.

We expect that {\it GEMS} will fulfill the high expectations and will motivate one or several
follow-up missions. A follow-up mission may feature:
\begin{itemize}
\item improved sensitivity over the 2 keV - 10 keV energy range combined with an 
improved energy resolution,
\item a broader energy bandpass with excellent sensitivity at lower ($<$2 keV) 
and/or higher ($>$ 10 keV) energies, 
\item the capability to do spectroscopic imaging polarimetry enabling the 
acquisition of 2-D maps of extended sources with spectroscopic and polarimetric information,  
\item a wide field of view (FoV) polarimeter with the possibility of measuring the polarization
of transient sources as for example Gamma-Ray Bursts.
\end{itemize}
In this paper, we focus on the possibility to measure the polarization of hard X-rays ($>$10 keV)
with narrow FoV instruments and wide FoV instruments. 
The soft gamma-ray telescope on board of the Japanese/US {\it ASTRO-H} mission 
(launch foreseen in 2013) will be able to do some hard X-ray Compton polarimetry using 
a combination of a collimator, Si pad detectors and CdTe pixel detectors \citep{Taka:08}.
With an effective area of $>$30 cm$^2$ for Compton scattered events, the soft gamma-ray telescope 
on board of {\it ASTRO-H} should be able to verify some of the theoretical predictions discussed in 
this paper if on-ground and in-orbit calibration measurements can be used to reduce the
systematic uncertainties below the level of the observed polarization effects.

In Section \ref{science} we summarize the 
science drivers for hard X-ray polarimetric observations. In Section \ref{designs} we  discuss 
four different experimental approaches suitable for small to mid size space missions, and 
present a comparative study of the performance of the different polarimeters based on 
Monte Carlo simulations. In Section \ref{discussion} we summarize the results of the 
previous sections and critically discuss which science objectives may be addressed 
with the experimental approaches discussed before.

The interested reader can consult \citep{Lei:97,Weis:06,Bell:10} for reviews of X-ray polarimetry
and for information about different X-ray polarimeters.
\section{Science drivers for hard X-ray polarimetry}
\label{science}
\subsection{Binary black holes}
Binary black hole (BBH) systems such as Cyg X-1 are among the brightest X-ray sources
in the sky and can reach flux levels during major flares similar to the flux 
from the Crab Nebula (see the review by \citet{Remi:06}).
The X-rays from a flat space Newtonian accretion disk are expected to be polarized 
owing to scatterings in the disk (e.g.\ \citep{Chan:60,Ange:69,Suny:85}).
\citet{Star:77}, \citet{Conn:80a}, and \citet{Conn:80b} showed that relativistic 
aberration and beaming, gravitational lensing, and gravito-magnetic 
frame-dragging result in an energy dependent polarization 
fraction and polarization degree. The energy dependence of the polarization
properties stems from the fact that higher energy photons originate closer
to the black hole and are affected more by relativistic effects than lower energy
photons. Recently, \citet{Dovc:04,Dovc:08} studied the polarization of
X-rays from accretion disks illuminated by a non-thermal X-ray source
``above'' the disk, as well as the observational consequences of 
different atmospheric depths. 
\citet{Agol:00} and \citet{Schn:09,Schn:10} included in their
calculation the effect of ``returning radiation'', X-rays which leave the 
accretion disk, hit it again after gravity deflects them around the black hole,
and are eventually scattered into direction of the observer.
At energies about the thermal peak (a few keV for typical BBH systems), 
the returning radiation can dominate the polarization signature, causing the 
polarization direction to swing from a horizontal orientation
at low energies (perpendicular 
to the rotation axis, as in the flat-space Newtonian limit discussed by 
Chandrasekhar) to vertical (parallel to the rotation axis) 
at higher energies.

\citet{Schn:09,Schn:10} studied the X-ray polarization in the thermal state
and in the hard/steep power law state (for definitions see \citep{Remi:06}). 
Spectropolarimetric observations of the thermal state make it
possible to test models of the radius dependent emissivity of the gas in the
accretion disk and the plunging region. If the X-ray emission follows a 
Novikov-Thorne zero-stress emissivity profile, the spectropolarimetric data can 
constrain the inclination of the inner accretion disk and the mass and spin of 
the black hole. Observations in the hard X-ray band have to cope with relatively low 
$>$10 keV fluxes in the thermal state. However, the hard X-ray polarization degree 
is expected to be much higher ($>$10\%) than in the 2-10 keV energy band 
($\sim$ 3 \%) owing to the dominance of returning radiation at high energies.
Owing to the soft thermal spectrum a good sensitivity in the 10 keV - 30 keV 
energy range will be important for these observations as a detection at
higher energies is unlikely. The hard X-ray observations are of great interest 
as they probe the innermost regions of the accretion flow constraining the properties of 
the accretion flow close to the black hole and the spin of the black hole.
Multi-epoch broadband spectropolarimetric observations can pin down the disk 
inclination and the black hole spin. The spin and inclination remain 
constant over very long times, while the disk, accretion rate, and coronal 
properties change on days-to-months times. High signal to noise ratio observations 
with good energy resolution may even make it possible to search for deviations 
from the General Theory of Relativity.

In the hard/steep power law (hard/SPL) state a hot corona comptonizes the disk emission.
Owing to the dominance of coronal emission, observations in the hard/SPL state
are not suited to constrain the properties of the central accretion flow nor the
spin of the black hole. However, hard X-ray observations will be key to probe the
properties of the corona. \citet{Schn:10} examined the spectropolarimetric 
signatures of various corona geometries. The emission properties depend on a large number
of parameters including the BH mass and spin, the BBH inclination, the accretion rate, and 
the coronal properties such as its homogeneity, the vertical scale height, the temperature, 
the optical depth, and the covering fraction. Constraining these parameters will require 
multiple spectropolarimetric observations at different flux levels. Although parameter 
degeneracies will likely render it impossible to pin down all these parameters, the X-ray 
polarimetry data will allow sensitive tests of corona models and their geometry. For example, a solid 
prediction resulting from uniform corona models is a high degree ($>$2-10\%) of polarization 
at $>$10 keV energies. A sensitive hard X-ray polarimeter may scrutinize the
BBH emission up to energies of 100 keV and above. A broad energy bandpass and 
a reasonable energy resolution (20\% or better) are thus important for exploring
BH coronas. Quasi-periodic oscillations (QPOs) are most pronounced at $>$6~keV 
energies and are thus
thought to originate in the hot corona \citep{Remi:06}. The comparison of the 
polarization signature with timing properties at different epochs should make it
possible to lead to constraints on the location, size, and coherence of the regions 
from which the QPOs originate \citep{Schn:10}.
\subsection{Neutron Stars, Pulsars, Pulsar Wind Nebulae, and Magnetars}
Magnetic neutron stars are expected to display a range of polarization 
phenomena in hard X-rays which have essential bearing on several aspects 
of the basic physics of radiation transfer in strongly magnetized plasmas. 

Isolated pulsars provide a neutron star source class that offers
exciting possibilities for polarimetric probes in the hard X-ray band.
Young pulsars possess high fields, and are bright in gamma-rays, which 
lead to their detection with the {\it Fermi} Gamma-Ray Space Telescope
\citep{Abdo:10b}.  The photons detected above 100 MeV are most likely
curvature radiation or synchrotron emission from tenuous pair plasmas in
the magnetospheres.  This contention applies also to the hard X-ray
emission in the 10 keV - 1 MeV window seen in famous pulsars like the
Crab and Vela.  Yet, because of pair cascading in both slot-gap 
(e.g.\ \citep{Musl:03}) and outer gap (e.g.\ \citep{Roma:96}) pictures for
gamma-ray pulsar emission, hard X-ray band radiation is more likely to
constitute a synchrotron signal, since its rate is very high when the
pairs produced have significant pitch angles. Synchrotron and curvature
emission possess similar spectral indices, whether coming from uncooled
electron populations, or strongly-cooled particles.  Hence another tool
is needed to discriminate between the two, and polarimetry can enable
such.
The degree of polarization from either process is high, and in uniform
fields couples to the spectral index for power-law electron
distributions according to Eq.~(\ref{eq1}) below.  In curved magnetospheric
fields, different observational perspectives relative to the local field
direction are sampled, depending on the pulse phase, and so the net
polarization signal can be lowered somewhat.  Pulse profiles can be used
as a diagnostic of typical magnetospheric locales using the gamma-ray
spectrum (e.g.\ \citep{Roma:10,Pier:10}).  Yet,
the orbital planes for accelerating charges that govern these two
radiation processes are orthogonal to each other. Accordingly, for very
confined emission regions, the Stokes vector for synchrotron emission is
perpendicular to that for curvature radiation. Superposing different
emission regions smears this discriminator somewhat, but the net product
is distinctive polarization angle (PA; vector on the sky) ``swings''
during the pulse period: the curvature pulse PA profile should display a
markedly different phase morphology from that for the synchrotron
process.  Hence, in conjunction with the light curve, measuring both the
polarization degree and angle as functions of phase and energy should
probe both the electron distribution shape, the spatial extent of the
emission region to some degree, and help decide which physical process
is dominating the signal in different X-ray bands. Ideal neutron stars
as candidates for such spectropolarimetric studies are the Crab, Vela
and PSR B1509-58 pulsars, because they are all very bright in 
X-rays, typically 20-50 mCrab in the range 50-500 keV, 
and their spectral index changes from 10 keV to 1 MeV.

Neutron stars with strong magnetic fields can exhibit cyclotron features occurring at the 
cyclotron resonance energy $E_c\equiv\hbar\omega_c\approx 11.6B_{12}(1+z)^{-1}$ keV, where
$B_{12}$ is the magnetic field B in units of $10^{12}$~G, and z is the gravitational redshift at the 
site of resonance (e.g., at or near the surface of a neutron star). Such features have now been
detected in the X-ray spectra of about fifteen accretion-powered pulsars in the energy 
range $\sim 15-50$ keV, indicating magnetic field strengths in the range $B_{12}\sim 1-5$
\citep{Cobu:02,Hein:04,Ghos:07,Naka:08}. 
Future hard X-ray studies may find similar features at higher energies in other pulsars, indicating 
higher magnetic fields.  
Near cyclotron resonance, oscillations in the \emph{amount} of linear X-ray polarization with pulse phase 
display a pronounced maximum in their amplitude \citep{Mesz:88}, which makes this energy range 
the optimal one for studies of such oscillations. The correlations between these oscillations in an 
accretion-powered pulsar and its pulse profile are one of the best known diagnostic probes of the beam 
shape of the pulsar. Detailed calculations have shown that for a pencil beam, oscillations in the 
polarization amount are expected to be \emph{out of phase} with the pulse (i.e., maximum of polarization 
at pulse minimum), while for a fan beam, polarization oscillations are expected to be \emph{in phase} 
with the pulse (i.e., maximum of polarization at pulse maximum) \citep{Mesz:88}. 
Moreover, there will be an energy-dependence to the polarization within a line feature,
and this will be different between the fundamental and higher harmonics due to photon spawning.
Measurements of these properties will provide diagnostics on viewing angle and effective optical depth.
Thus hard X-ray
polarization studies near the cyclotron resonance energies of accretion-powered pulsars have great 
potential for being pioneering probes into the beam shapes of these pulsars, which will constrain 
models of accretion flow to the magnetic poles of such pulsars.
Bright accreting pulsars like 4U 0115+63 can exhibit 10 keV - 50 keV fluxes of $\sim$1/4 Crab \citep{Cobu:02}.
The particular source 4U 0115+63 is a good target as five cyclotron resonance features were 
observed up to energies of 50 keV. At higher energies the spectrum runs out of statistics \citep{Hein:00}.

Birefringence of the vacuum in a strong magnetic field is a prediction of fundamental physics which may
be amenable to direct demonstration through X-ray polarization studies of accretion-powered pulsars.
The crucial energy in this context is that of vacuum resonance $\hbar\omega_0\approx 13B_{12}^{-1}
n_{e,22}^{1/2}$ keV, where vacuum birefringence effects cancel the plasma effects, and the 
\emph{direction} of polarization is expected to rotate by 90 degrees due to \emph{mode conversion}. 
Here, $n_{e,22}$ is the electron density in units of $10^{22}$ cm$^{-3}$. Mode conversion is a 
phenomenon which occurs through an interplay between (a) the photospheres corresponding to the ordinary 
and extraordinary modes of electromagnetic wave propagation in magnetized birefringent plasmas, and 
(b) the above vacuum resonance point (\citep{Lai:03a,Lai:03b} and references therein). In accretion-powered 
pulsars with magnetic fields in the range $10^{11}-10^{12}$ G and at accretion rates corresponding to 
electron densities in the range $10^{22}-10^{23}$ cm$^{-3}$, the vacuum resonance energy occurs in the 
hard X-ray band, underscoring the importance of hard X-ray polarization studies in exploring this 
phenomenon. Indeed, a measurement of the above rotation of polarization would be the first observational 
confirmation of the fundamental idea of vacuum birefringence.
Bright accretion powered pulsars like Her X-1 and Cen X-3 exhibit hard X-ray 
fluxes on the order of 50 mCrab \citep{Luto:09}.                  
             
For magnetars, which are believed to be isolated neutron stars with super-strong magnetic fields in the
range $10^{14}-10^{15}$ G, the above ideas of vacuum resonance lead to a completely different 
prediction \citep{vanA:06}. The direction of X-ray polarization in these objects is expected
to remain \emph{unchanged} as one passes through the vacuum resonance point. The vacuum resonance 
energy is somewhat more difficult to estimate for these objects due to uncertainties in the plasma
densities at the sites of their X-ray emission, but it is clear from the above estimate that it will be
in the hard X-ray / soft $\gamma$-ray band for densities $n_e\sim 10^{17}-10^{20}$ cm$^{-3}$. A 
demonstration of this differential behavior in polarization rotation between known accretion-powered 
pulsars and suggested magnetars would be a strong indication of a different role of vacuum resonance 
in the latter, and so a probe of exotic phenomena in super-strong magnetic fields. Particularly suitable 
for such studies would be the soft gamma repeaters (SGRs). 
The SGR 1806-20 normally exhibits hard X-ray fluxes of 0.2 mCrab \citep{Espo:07}. 
During the giant outburst of December 2004, this SGR exhibited hard X-ray fluxes 
on the order of the flux from the Crab Nebula. The $>$80 keV flux integrated over 400 sec 
was 2.6$\times 10^{-4}$ erg cm$^{-2}$ \citep{Mere:05}.

In the lower density magnetospheres of magnetars different polarizing
influences can be found.  Observationally, this is probably mediated by
magnetic Compton scattering in the very strong fields, which is
extremely efficient, more so than curvature emission or bremsstrahlung.
The manifestation of Compton scattering comes in two varieties in two
different wavebands.  In the classical X-ray band, anomalous X-ray
pulsars and soft gamma repeaters both exhibit steep X-ray power-law
tails appended to the thermal peak (e.g.\ \citep{Pern:01}). These have
been modeled by repeated Compton scattering by mildly-relativistic
electrons (e.g.\ \citep{Lyut:06,Nobi:08}) in
collisions well below the resonance in the cross section at the
cyclotron energy.  The collision rates are strongly-dependent on the
angle the photon momenta make to the local magnetic field, and to their
polarization state.  Hence, phase-dependent polarization observations of
these tails below around 10 keV should probe both the locale of the
emission (equatorial versus near the magnetic pole) and the observer's
viewing angle.  Quiescent magnetar signals also exhibit hard X-ray tails
extending up to around 150 keV \citep{Kuip:06,Goet:06,Hart:08}.  
These probably are generated by impulsive
inverse Compton scattering of soft thermal X-rays by ultra-relativistic
electrons in the strong magnetar fields (e.g.\ \citep{Bari:07,Nobi:08}), 
sampling the collisional cross section at or near
the cyclotron resonance.  The up-scattering spectra are strongly
polarized in an energy-dependent fashion \citep{Bari:07}.
Phase-dependent spectropolarimetry can confirm the postulate of the
action of resonant Compton scattering in generating these hard tails,
and again provide insights into the magnetospheric locale of the
emission region.  A number of magnetars including 4U 0142+61, 
RXS J1708-4009, SGR 1806-20 and SGR 0501+4516 are amenable to such
polarimetric investigations with a polarimeter which can measure 
10\% polarization fractions for sources at the detection thresholds 
of {\it RXTE} and {\it INTEGRAL}. Quiescent fluxes from these sources 
are generally at the 5--20 mCrab level in the 20--200 keV range.

The hard X-ray tails do not extend to much higher energies, as inferred
from very constraining upper limits at $>500$ keV from the {\it COMPTEL}
instrument on the Compton Gamma-Ray Observatory (e.g.\ \citep{Hart:08}). 
The {\it Fermi} Large Area Telescope (LAT) can be used to derive additional 
constraints on the high-energy properties of the energy spectra, but 
so far dedicated studies have not yet been published.
If the maximum energies correspond to kinematic limits in the
scattering process, then turnovers in the 200-500 keV band
should have characteristic energies independent of the photon
polarization.  However, in the strong fields not far from the magnetar
surface, photon splitting $\gamma\to\gamma\gamma$ can act as an
attenuation mechanism that may forge a spectral turnover at these
energies.  Splitting can only occur in strong magnetic fields 
(e.g.\ \citep{Bari:08}), and is an exotic prediction of QED.  Its rate is strongly
dependent on the polarization of the photon, so that measurements of
quasi-exponential turnovers whose energies depend markedly on the photon
polarization would be a telltale sign of the action of splitting.
Detecting such a signature would be an important confirmation of a 
theoretical prediction, since magnetic photon splitting has not been 
confirmed in the laboratory.
Through the dependence of the splitting rate on the angle between
the photon momentum and the magnetic field line direction and the strength 
of the local field, phase-dependent polarimetric observations would facilitate determination
of the altitude and co-latitude of the emission region.  Strong flares in
SGRs possess quasi-thermal spectra with effective temperatures in the
15-40 keV range.  It is believed that these emanate from fireballs 
(e.g.\ \citep{Thom:96}), whose opacity is largely controlled by
magnetic Compton scattering.  At such energies, photon splitting should
also be active in modifying the higher frequency portion of the
spectrum.  Accordingly, such bright outbursts, and the even stronger
giant flares from SGRs (around 8-10 orders of magnitude brighter than 
magnetar quiescent emission), if detected by an all-sky monitor,
offer excellent opportunities to probe the
radiation physics and emission region conditions in this subset of
magnetars.
\subsection{Active Galactic Nuclei}
Active Galactic Nuclei (AGNs) emit thermally in the hard UV band
with the peak of the thermal emission falling into the 30-100 eV range.
At X-rays, a hard power law emission component is observed with 
photon indices $\Gamma\,=$ 1.5-2 \citep{Nand:91,Mush:93}.
The power law emission is thought to originate from the corona through the 
Comptonization of thermal photons by electrons of a temperature 
$T_{\rm c}\,\sim$100 keV \citep{Haar:93}.
The polarization of Comptonized disk emission was studied by 
\citet{Suny:85} and \citet{Schn:10}. The coronal emission is expected to be 
polarized to a higher degree (typical values of $\sim 8$\%) than in the case of BBHs
(typical values of $\sim 2$\%). 
Owing to the large number of interactions required to increase the energy of the thermally 
emitted disk photons from a fraction of a keV to $>$10 keV, photon paths parallel to the 
plane of the accretion disk and the corona are more common than in the case of the 
photons from BBH coronae.
\citet{Suny:85} and \citet{Schn:10} find that the spectropolarimetric data 
can be used to constrain the inclination of the AGN. 
In the case that the inclination of the disk is known
(e.g.\ from the blue edge of the broad iron line \citep{Reyn:03}),
the X-ray spectropolarimetric data can be used to constrain details
of the coronal model such as the number of clumps and their over-density.
X-ray polarimetry studies of AGNs have to cope with low fluxes:
Type 1 Seyfert AGNs like NGC 4151 and NGC 5548 have 
X-ray fluxes on the order of 1 mCrab \citep{Piro:98,Piro:00}.
Furthermore, the interpretation of the data may be complicated owing
to a contamination of the emission by the non-thermal emission 
from a jet.
\subsection{Jets from Active Galactic Nuclei}
The highly relativistic jets (collimated outflows) from accreting supermassive black holes 
at the center of galaxies are sources of electromagnetic continuum radiation all the way 
from the radio band to the TeV gamma-ray band. Of particular interest for the 
test of jet formation models are blazars observations. Blazars are accreting 
supermassive black hole systems with one jet aligned with the line of sight. 
The relativistic motion of the emitting plasma amplifies the jet emission along the 
direction of motion and makes blazars some of the brightest extragalactic objects in the Universe. 
The spectral energy distribution of the continuum emission shows a low-energy peak 
and a high-energy peak. According to the standard paradigm a population of relativistic 
electrons emits the low-energy and high-energy components as synchrotron and inverse Compton emission, respectively.
Magnetic fields are thought to be an important jet component. They may even 
dominate the jet energy density. Furthermore, the magnetic fields are dynamically important and are believed to stabilize jets 
on spatial scales between a few light hours and thousands of light years.
\begin{figure}[tb]
\begin{center}
\includegraphics[width=3.5in]{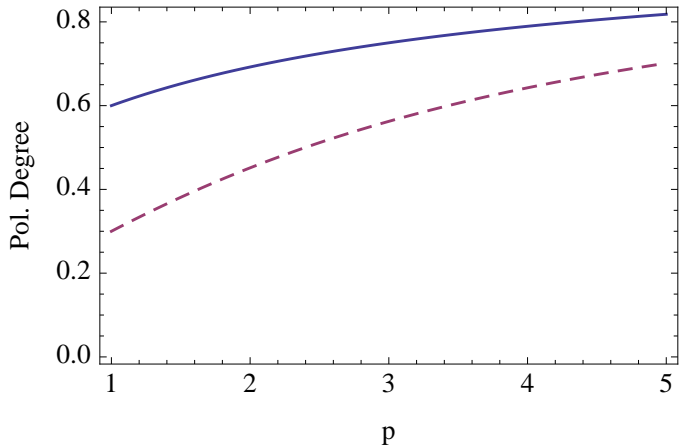}
\caption{\label{yp} Polarization degree of the synchrotron emission (solid line) and the
inverse Compton emission (dashed line) as function of the electron spectral index
in SSC models of the continuum emission from blazars. The curves assume a uniform 
magnetic field, and $\sin{\zeta}\,=1$ for the angle $\zeta$ between the magnetic field
direction and the propagation direction of the SSC photons in the jet frame.}
\end{center}
\end{figure}

For low-energy peaked blazars (i.e.\ flat spectrum radio quasars) the low-energy emission component 
peaks at infrared/optical wavelengths, and the high-energy emission component peaks in the 
MeV gamma-ray regime. The X-ray to gamma-ray emission of such objects is believed to be 
inverse Compton emission. Multiwavelength observations of the polarization degree and polarization 
direction have the potential to elucidate how the hard X-rays are emitted. In the Synchrotron 
Self-Compton (SSC) scenario the X-rays come from inverse Compton interactions of electrons 
with co-spatially emitted synchrotron photons, and one expects that the X-ray polarization tracks the polarization of the synchrotron 
emission at longer (radio, infrared, and optical) wavelengths \citep{Pout:94,Celo:94,McNa:09}. 
The optical emission of blazars shows high polarization degrees  
\citep{Ange:80,Scar:97} close to the theoretical expectation for a power law electron distribution
with significant pitch angles (exceeding 1/$\gamma$, with $\gamma$ the electron Lorenz factor)  
in a uniform magnetic field:
\begin{equation}
P_{\rm S}\,=\,\frac{p+1}{p+7/3}
\label{eq1}
\end{equation}
where p is the spectral index of the electron energy spectrum. Poutanen (1994) gives a simple 
expression for the polarization degree of the SSC emission $P_{\rm SSC}$ as 
function of $P_{\rm S}$ and $p$:
\begin{equation}
P_{\rm SSC}\,=\,P_{\rm S} \frac{(p+1)(p+3)}{p^2+4p+11} \sin{^2\zeta}
\end{equation}
with $\zeta$ being the angle between the magnetic field direction 
and the SSC emission in the jet frame. Note that $\sin{\zeta}\sim 1$ is
quite likely even if the angle between the jet axis and the line of 
sight is only a few degrees owing to relativistic aberration. 
Figure \ref{yp} shows the synchrotron and SSC polarization 
degrees as function of the electron spectral index. 
{\it GEMS} should be able to decide between the SSC scenario and alternative 
external Compton (EC) models in which the target photons are provided by the accretion disk, 
the broad line region, from an outer slow jet sheet, or from other upstream or downstream 
regions of the jet. 
Whereas the synchrotron and inverse Compton emission 
should show similar polarization degrees and identical polarization directions
in the SSC scenario, EC models predict much lower polarization fractions ($<$5\%)
and polarization directions that depend on the seed photon source and the alignment of
the jet to the line of sight \citep{McNa:09}. High signal-to-noise time resolved broadband 
spectropolarimetric observations (e.g.\ 0.1 keV-100 keV) would make it possible to sample 
the time evolution of the energy dependent flux, 
polarization fraction, and polarization direction of several spectral components. 
The observations of the polarization properties would afford the possibility to decide if
different spectral components are emitted co-spatially (similar polarization 
degrees and directions) or not (different polarization degrees and directions). 
If SSC models hold, polarimetric multiwavelength observations 
may make it possible to identify spectral parts (e.g.\ in the optical and X-ray bands)
which are emitted by electrons of the same population.

For high-energy peaked blazars (i.e.\ BL Lac objects), the low-energy emission component peaks 
in the X-ray band, and the high-energy emission component peaks in the GeV to TeV 
gamma-ray regime. In the case of ``extreme synchrotron blazars'' like Mrk 421, Mrk 501,
and 1ES 1959+650 the low-energy synchrotron component extends from the radio band 
all the way to hard ($\sim$100 keV) X-ray band. These objects exhibit correlated X-ray and 
gamma-ray flares on time scales of minutes. The fast flares indicate 
that the X-ray/TeV gamma-ray emission originates close to the black hole where the jet energy 
density is highest and the jet cross section is smallest. The observations of the polarization 
of synchrotron  X-rays might thus allow us to probe the structure of the magnetic field 
close to the base of the jet. Of particular interest are observations of high degrees 
of polarization close to the theoretical maximum indicating the presence of a uniform
magnetic field. Broadband spectropolarimetric observations 
(e.g.\ 0.1 keV-100 keV) of curved energy spectra would make it possible to determine the 
polarization degree at different energies and for different electron spectral indices
and thus to test if the polarization fraction depends on the electron spectral index
as predicted by Equ.\ (\ref{eq1}).

Recently, optical polarization swings by 180$^{\circ}$ in temporal 
coincidence with X-ray and gamma-ray flares were reported
\citep{Mars:08,Abdo:10a}. If jets are threaded and accelerated by helical magnetic fields
polarization swings may be produced by the field moving through a stationary shock 
feature \citep{Mars:08}. If the explanation is correct it should be possible to
observe such polarization swings in the soft and hard X-ray bands and to confirm 
the presence of a helical magnetic field structure. 
If the global field morphology within the jet is quite turbulent, there may be 
significant depolarization, and the temporal variations in the degree and angle 
of polarization may vary in a more chaotic manner.

For all three blazar science topics described above hard X-ray observations 
have one distinct advantage over soft X-ray observations: the electrons responsible for
the hard X-ray emission cool faster than the electrons responsible for
the soft X-ray emission. Hard X-ray observations can thus resolve time variable
phenomena on shorter time scales -- if the signal to noise ratio is sufficiently high. 
Bright blazars like Mrk 421, Mrk 501, and PKS 2155-304 can reach X-ray fluxes between 
10 mCrab and 100 mCrab.   
\subsection{Gamma-Ray Bursts}
GRBs are brief flashes of gamma-rays from random locations in the sky. 
For a short time period, the GRB becomes the most luminous object in the Universe.
The GRB emission is thought to originate in highly relativistic jets 
powered by hypernovae (explosions of massive stars) \citep{Woos:93,Pacs:98}
or by coalescing neutron stars or neutron star black hole mergers (see the references in \citep{Gehr:05}).
Gamma-ray opacity arguments constrain the jet bulk Lorentz factors to values $\gg$100 \citep{Rees:94}.
The jet launching mechanism is still highly uncertain. In the case of long GRBs ($>$2 s) 
a newly formed spinning black hole inside a collapsing star may launch the jet through a 
combination of thermal pressure, electric fields generated by the black hole spinning 
in the magnetic field of the collapsing star, inertia of the stellar material, 
magnetic pressure, and magnetic stresses (see \citep{McKi:06} for 
a comprehensive discussion).
\citet{Cobu:03} reported a strong polarization of GRB021206, however, independent analyses
did not confirm the result \citep{Rutl:04}. \citet{Kale:07,McGl:07,Goet:09} 
reported on {\it INTEGRAL} observations of GRB 041219a which were consistent 
with a high degree of polarization but did not support the polarization with a
high statistical significance.  

X-ray polarimetry has the potential to make smoking gun measurements with regards to the 
mechanism responsible for the prompt emission and the structure of GRB jets 
\citep{Gran:03,Eich:03,Lyut:03,Waxm:03,Ross:04}. 
If the jets are threaded by uniform magnetic fields and the prompt X-ray/gamma-ray emission 
is synchrotron emission a GRB polarimeter would measure high polarization fractions. 
If GRB jets are threaded by helical magnetic fields that accelerate and confine the jets 
the polarization direction may exhibit a continuous swing as the helical field moves 
through a standing shock feature (e.g.\ as the jet exits the star) or as the shock 
moves relative to the particle accelerating shock. 
This would then provide a hard X-ray analog for GRBs of the polarization swing observed in
the optical for the blazar 3C 279 \citep{Abdo:10a}.
  
Whereas a wide FoV hard X-ray experiment would be able to autonomously detect GRBs 
a narrow FoV experiment would have to rely on the detection of GRBs by other experiments to perform
rapid follow-up observations.
The Swift BAT localizes GRBs with an accuracy of 1-4 arcmin 
within 20 sec after the start of the event \citep{Bart:05}. Subsequent observations 
with the Swift X-Ray Telescope (XRT) improve the localization accuracy to a few arcsec 
a few minutes after the start of the event. The localization errors of the {\it Fermi} 
GRB detectors are of the order of a few degrees. Autonomous detection 
of GRBs requires a wide FoV instrument covering a substantial 
fraction of the sky. 

\begin{figure}[tb]
\begin{center}
\includegraphics[width=3.5in]{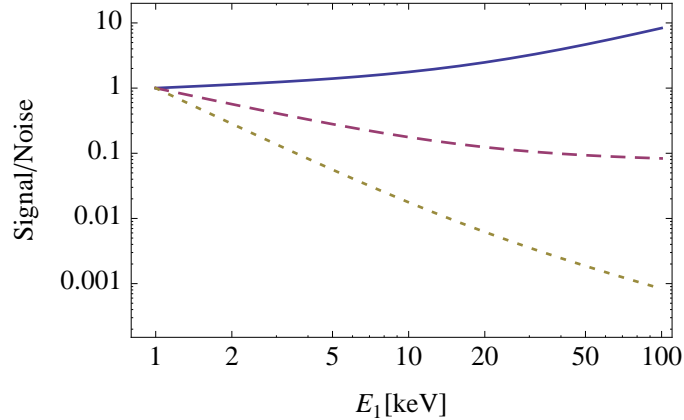}
\caption{\label{GRBsens} Signal to noise ratio (SNR) for the detection of GRBs 
with a wide field of view instrument as a function of the threshold energy $E_1$. 
The SNR was computed for three different GRB photon indices ($dN/dE\,\propto$
$E^{-\Gamma}$): $\Gamma\,=$ 1 (solid line), 
$\Gamma\,=$ 2 (dashed line), and $\Gamma\,=$ 3 (dotted line).
The calculations assumed a SNR limited by the noise of the diffuse 
cosmic X-ray background (CXB) \citep{Ajel:08} and instruments with a 
uniform sensitivity from $E_1$ ... $E_2\,=$ 4~$E_1$. 
All curves were normalized to unity at $E_1\,=$ 1 keV.}
\end{center}
\end{figure}

In the 0.3 keV - 150 keV energy range the energy spectra of the prompt GRB emission 
can be described with power laws. The photon indices $\Gamma$ 
(dN/dE $\propto$ E$^{-\Gamma}$) vary over a wide range from $\approx$1 to 
$\approx$3 \citep{OBri:06}. For a space borne mission in Low Earth Orbit 
the $<$ $\sim$150 keV background usually is dominated by the diffuse 
extragalactic background radiation. 
Figure \ref{GRBsens} shows the signal to noise ratio as function of the threshold energy
$E_1$, assuming an instrument with a bandpass from $E_1$ to $E_2\,=\,4\,E_1$.
The graphs assume a GRB photon flux proportional to 
$dN/dE\,\propto$ $E^{-\Gamma}$ with $\Gamma\,=$ 1, 2 and 3, and 
the diffuse extragalactic background radiation model of \citep{Ajel:08}.
For the hardest GRBs with $\Gamma\,=$ 1 the signal to noise ratio improves with 
$E_1$ showing that a large effective area and a wide energy bandpass are most important 
for the detection of these bursts. For softer bursts a low energy 
threshold is essential for a sensitive observatory.

To summarize: a GRB polarimetry mission may have a rather narrow FoV
as long as other GRB missions are still operational and can distribute rapid GRB 
alerts. For all but the hardest GRBs a low energy threshold is crucial for 
detecting the GRBs. An instrument with a broad energy coverage has the 
obvious advantage to be able to measure the energy dependence of the 
polarization properties.
The first catalog of GRBs detected with the Swift BAT in 2 1/2 years 
contains 237 GRBs with 15-150 keV fluences between 6$\times 10^{-9}$ erg cm$^{-2}$ 
and 4$\times 10^{-5}$ erg cm$^{-2}$ \citep{Saka:08}.
\subsection{Solar and Stellar Hard X-ray Flares}
Solar flares are powerful events releasing up to 10$^{33}$ erg of which
a substantial fraction goes into accelerating ions.
Exceptional solar flares can accelerate ions up to energies of several 
10 GeV and electrons up to energies of several 100 MeV \citep{Lin:76}.
Solar flares emit $\sim$100 keV photons as Bremsstrahlung. 
Even though the polarization of the X-rays may be influenced by Compton
scatterings of the X-rays before they leave the Sun, the hard X-ray polarization 
measurements have a unique potential to constrain the electron beaming 
and the orientation of the magnetic field loops with respect to the
line of sight \citep{Korc:67,Elwe:68,Emsl:08}.
Whereas X-ray energy spectra are largely independent of the solid angle distribution
of the momenta of the emitting electrons, the expected polarization degree varies from a few 
percent for an approximately isotropic electron distribution to 20\%-25\% 
for highly beamed distributions \citep{Bai:78,Leac:83,Chan:88}.
The X-ray continuum may be contaminated by thermal emission at lower ($<$50 keV) 
energies, but above 1 MeV the gamma-ray emission mostly consists of  
unpolarized nuclear line emission. 
The best energy range for polarimetric observations is thus
the window between 50 keV and 1 MeV. First tentative detections of the X-ray 
polarization of solar flares were reported by \citet{Bogo:03,Bogg:06,Suar:06,Zhit:06}.
During a solar flare, the direction of the polarization vector relative to the magnetic
field direction will depend on the pitch angle distribution of the emitting electrons and on
the viewing angle. Firm conclusions about the direction distribution of the emitting
electrons can thus require the measurement of the polarization of many solar flares 
and by exploring the dependence of the polarization properties on the viewing angle.
The rate of solar flares depends on the solar activity. 
\citet{Cros:93} report differences of the rates of solar flares 
of certain peak fluxes of between 1 and 2 orders of magnitude between
solar cycle maximum and solar cycle minimum.
Solar flares can be very bright with $>$20 keV peak fluxes exceeding
1000 photons cm$^{-2}$ s$^{-1}$ keV$^{-1}$ \citep{Cros:93}.

In principle, it should also be possible to detect hard X-ray flares from non-thermal
electron populations from other stars. So far, hard X-ray studies of stellar flares 
were limited by the sensitivity of the detectors. All the hard X-ray emission recorded 
so far (e.g.\ by the BeppoSax satellite, \citep{Fran:99}) 
could be explained as thermal emission \citep{Gued:04}. 
Polarimetric studies of the hard X-ray emission from stars 
are beyond the sensitivity of the instruments discussed in this paper.
\subsection{Search for Lorentz Invariance Violation}
For the last two decades theoretical studies and experimental searches of Lorentz 
Invariance Violation (LIV) have received a lot of attention 
(see the reviews by \citep{Matt:05,Jaco:06,Will:06,Libe:09}).
On general grounds one expects that the two fundamental theories of our time
the General Theory of Relativity and the Standard Model of Particle Physics
can be unified at the Planck energy scale. Under certain conditions deviations 
from the two theories may be observable even at much lower energies, 
e.g.\ if effects such as a tiny difference between the propagation 
speed of orthogonally polarized photons accumulate over cosmological distances to become measurable \citep{Coll:97,Coll:98}.

Possible consequences of LIV are energy and helicity dependent photon 
propagation velocities. The energy dependence can be constrained by 
recording the arrival times of photons of different energies emitted 
by distant objects at approximately the same time \citep{Amel:98}, 
e.g.\ during a Gamma-Ray Burst \citep{Abdo:09}
or a flare of an Active Galactic Nucleus \citep{Ahar:08}. 
The energy and helicity dependence can be constrained by measuring 
how the polarization direction of an X-ray beam of cosmological 
origin changes as function of energy \citep{Gamb:99}.
Since linear polarization is a superposition of two monochromatic waves 
with opposite circular polarizations, a helicity dependent speed of light
would lead to a rotation of the polarization direction along the photons' path.
As the deviation of the speed of light from its value at low photon energies 
would depend on energy, the propagation through space should lead to a 
frequency dependent polarization swing superimposed on the source inherent
variation of the polarization properties.
Depending on the dispersion relations for the different photon helicities, 
observations over a given frequency range could lead to three different outcomes:
(i) the effect could be too small to lead to observable consequences, 
(ii) it could have just the right magnitude to lead to a swing of the 
polarization direction in the covered frequency range, and (iii) it could
be so large that even observations over a narrow energy band (given by the 
energy resolution of the telescope) would average over several  
polarization swings and would lead to a zero net-polarization.
The observation of a linearly polarized signal from a cosmological 
source can thus be used to set an upper limit on the magnitude of the LIV effect
(see \citep{Fan:07} and references therein). 

If the photon velocity depends on the photon helicity, birefringence 
measurements tend to yield more sensitive tests of the underlying 
models than time dispersion measurements. If the source of the
photons is at distance $L$ and photons exhibit a group velocity
difference $\Delta\,v_{\rm g}$ the measured time dispersion is 
\begin{equation}
\Delta t \,\approx\,\frac{L \Delta v_{\rm g}}{c^2} 
\end{equation}
The detection of an extremely fast (1 $\mu$s) burst of X-rays from 
a GRB at a light travel time of $L\,=$ 1~Gpc would allow us to probe 
group velocity differences down to
\begin{equation}
\frac{\Delta v_{\rm g}}{c} \,\approx\,
\Delta t\,(L/c)^{-1}\,\approx\,
10^{-23}\,
\frac{\Delta t}{\rm 1 \, \mu s}\left(\frac{L}{\rm 1\,Gpc}\right)^{-1} 
\label{disp}
\end{equation}
The same group velocity difference would lead to a phase difference
\begin{equation}
\Delta \phi \,\approx\,\frac{L \nu \Delta v_{\rm g}}{c^2} 
\end{equation}
where $\nu$ is the (mean) frequency of the observed photons. 
A measurement of the phase difference with an accuracy of $\Delta\phi\,=$ 
10$^{\circ}$ would allow us to access group velocity differences down to
\begin{equation}
\frac{\Delta v_{\rm g}}{c} \,\approx\,
\Delta\phi (L/\lambda)^{-1} \,\approx\, 10^{-37}\,
\frac{\Delta \phi}{\rm 10^{\circ}}
\left(\frac{E_{\gamma}}{\rm 100\,keV}\right)^{-1} 
\left(\frac{L}{\rm 1\,Gpc}\right)^{-1} 
\label{LIVeq}
\end{equation}
where $E_{\gamma}$ and $\lambda$ are the mean energy and wavelength of the observed 
photons. Comparison of Eqs.\ (\ref{disp}) and (\ref{LIVeq}) demonstrates the 
power of polarimetric observations.

Equation (\ref{LIVeq}) shows that deviations of the speed of light
can be measured with an accuracy proportional to $(L/\lambda)^{-1}$ $\propto\,E_{\gamma}^{-1}$. 
Measurements at higher energies and shorter wavelengths thus lead to better constraints.
LIV is thought to be a high-energy phenomenon; the deviation of the
speed of light is expected to increase with energy.
Using an expansion in powers of $E_{\rm \gamma}$
\begin{equation}
\frac{\Delta v_{\rm g}}{c}\,=\,\sum_{n=1}^{\infty} \eta^{(n)} \left(E_{\gamma}/E_{\rm Pl}\right)^{\,\,n}
\end{equation}
with $E_{\rm Pl}$ the Planck energy, we infer that the accuracies of the birefringence 
constraints on the coefficients $\eta^{(n)}$ scale as $E_{\gamma}^{-(n+1)}$.
Presently, the best constraints on the LIV coefficients come from the observations 
of polarized optical/UV emission from a GRB afterglow at redshift $z\,=$ 1.255 
at a frequency of 5$\times$10$^{14}$ Hz \citep{Fan:07}. Polarimetric hard X-ray observations
($E_{\gamma}\,\sim$100 keV) of a GRB or an AGN at a similar redshift would improve on these 
limits by 9 orders of magnitudes for $n\,=$ 1.
For $n\,=$ 2 hard X-ray polarimetry would lead to an even larger improvement by 
14 orders of magnitudes.
Note that the concept of birefringence involves anisotropy.
General constraints on polarization dependent LIV thus require polarimetric observations
of a sample of GRBs or AGNs from different directions \citep{Kost:09}. 
\section{Hard X-ray polarimeters: four technical approaches and their performance}
\label{designs}
\subsection{General considerations}
The linear polarization of photons with energies between $\sim$10 keV 
and several MeV can best be measured based on the Compton effect. 
Such measurements take advantage of the 
fact that linearly polarized photons are preferentially scattered into
a direction perpendicular to the electric field orientation.
Combining a low-Z scatterer with a high-Z absorber leads to a high probability 
for a Compton interaction in the low-Z material followed by a photoeffect absorption 
in the high-Z material. The polarization can be measured by histogramming the distribution 
of azimuthal scattering angles and fitting a sinusoidal modulation to the histogram. 
The modulation of the recorded azimuthal angles depends on the energy 
of the incident photons, the polar scattering angle, and the design 
of the detector assembly.
\begin{figure}[tb]
\begin{center}
\includegraphics[width=3.5in]{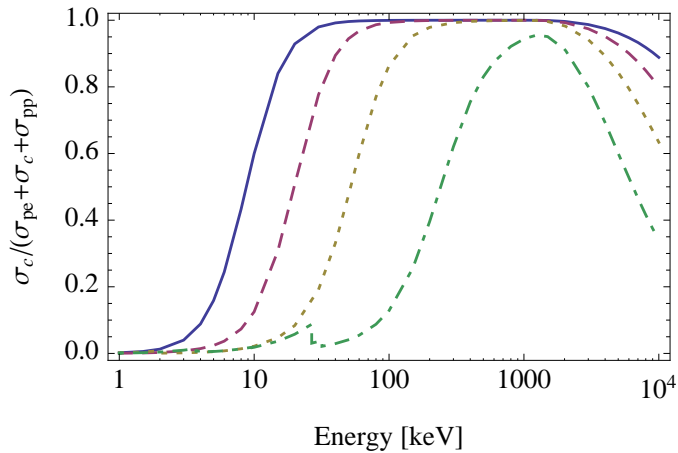}
\caption{\label{sigmaCS} Compton scattering efficiency 
as function of energy for  Li, C, Si, and CZT (from left to right). 
The efficiency is defined here as the Compton scattering cross 
section divided by the total cross section. The cross section 
data are from \citep{Berg:10}. 
}
\end{center}
\end{figure}
At low photon energies the dominance of photoeffect interactions 
over Compton interactions results in a ``low energy threshold'' 
of a polarimeter that depends on the atomic number $Z$ of the Compton scatterer.
Figure \ref{sigmaCS} shows the ``efficiency for Compton scattering'' 
for different materials where the efficiency $\epsilon$ is defined as the ratio 
of the scattering cross section (including coherent scattering) 
and the total cross section. 
If we define a threshold energy by the condition $\epsilon>0.1$ we get 
threshold energies of $E_{0.1}\,=$ 4.2 keV for Li, $E_{0.1}\,=$ 9.0 keV for C, 
$E_{0.1}\,=$ 21.5 keV for Si, and $E_{0.1}\,=$ 86.1 keV for the semiconductor 
compound $\rm Cd_{0.9}Zn_{0.1}Te$.
The 50\%-efficiency points are reached at the energies $E_{0.5}\,=$ 8.8 keV for Li,
$E_{0.5}\,=$ 19.8 keV for C, $E_{0.5}\,=$ 52.1 keV for Si, and
$E_{0.5}\,=$ 246.4 keV for $\rm Cd_{0.9}Zn_{0.1}Te$.

\begin{figure}[tb]
\begin{center}
\includegraphics[width=3.5in]{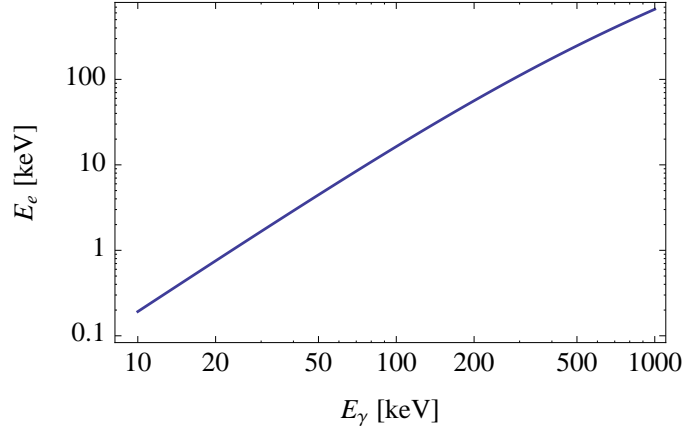}
\caption{\label{ComptonEl} The energy of the Compton electron $E_{\rm e}$
as function of the energy of the primary photon $E_{\gamma}$ for a polar
scattering angle of 90$^{\circ}$. 
}
\end{center}
\end{figure}
The background can be reduced by requesting the temporal coincidence between
a hit detected in the low-Z scatterer and in the high-Z absorber.
In this case the energy threshold of the low-Z detector is likely to determine 
the energy threshold of the polarimeter because photons with energies of a few 
10 keV ($\ll m_{\rm e}c^2$) deposit only a small fraction of their energy in 
the first interaction. Assuming a scattering angle of 90$^{\circ}$ and neglecting the
binding energy of the electron, the energy of the Compton electron as function 
of the energy of the primary photon $E_{\gamma}$ is (see Fig.\ \ref{ComptonEl}):
\begin{equation}
E_{\rm e}\,=\,\frac{E_{\gamma}^2}{m_{\rm e}c^2+E_{\gamma}}
\end{equation}
The Compton electron receives an energy of $E_{\rm e}\,=$ 0.2 keV, 1.2 keV, 
4.5 keV, and 16 keV for $E_{\gamma}\,=$ 10 keV, 25 keV, 50 keV, and 100 keV, respectively.

As mentioned above the information about the polarization of the gamma-ray beam 
is encoded in the azimuthal scattering angle distribution. The sensitivity of
a polarimeter depends on the modulation factor
\begin{equation}
\mu\,=\,\frac{n_{\rm max}-n_{\rm min}}{n_{\rm max}+n_{\rm min}}
\label{modF}
\end{equation}
where $n_{\rm max}$ and $n_{\rm min}$ refer to the maximum and minimum 
counts in a azimuthal scattering angle histogram.

\begin{figure}[tb]
\begin{center}
\hspace*{-0.5cm}
\begin{minipage}[htb]{6.5cm}
\includegraphics[width=6.5cm]{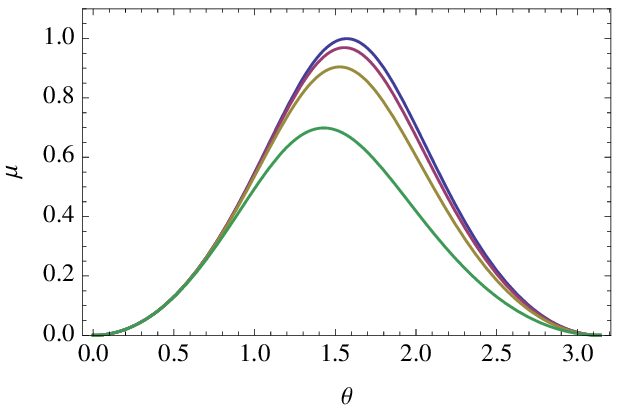}
\end{minipage}
\begin{minipage}[htb]{6.5cm}
\vspace*{-0.3cm}
\includegraphics[width=6.5cm]{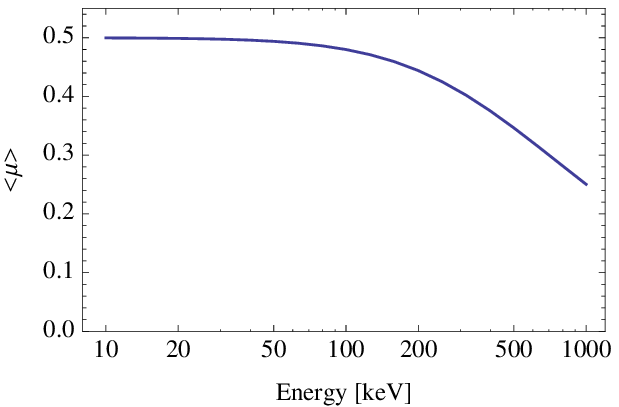}
\end{minipage} 
\caption{\label{muFig} The left panel shows the modulation factor as function of the
polar scattering angle for photon energies of 
10, 100, 200, and 500 keV (from top to bottom). The right panel shows the 
modulation factor averaged over the solid angle and the Klein-Nishina cross section.
}
\end{center}
\end{figure}
The modulation factor of Compton scattered radiation can be computed based on the 
Klein-Nishina cross section \citep{Eva:55}:
\begin{equation}
\frac{d\sigma}{d\Omega}\,=\,
\frac{r_0^2}{2}
\frac{k_1^2}{k_0^2}
\left[
\frac{k_0}{k_1}+\frac{k_1}{k_0}-2 {\rm sin}^2 \theta {\rm cos}^2 \eta
\right]
\label{KNequ1}
\end{equation}
with $r_0$ the classical electron radius, {\bf k}$_0$ and {\bf k}$_1$ the
wave-vectors before and after scattering, $\theta$ the scattering angle
(the angle between {\bf k}$_0$ and {\bf k}$_1$), and $\eta$
the angle between the electric vector of the incident photon 
and the scattering plane. The equation for the modulation factor reads:
\begin{equation}
\mu(\theta) \,=
\frac{
\left(\frac{d\sigma}{d\Omega}\right)_{\eta=\pi/2}-\left(\frac{d\sigma}{d\Omega}\right)_{\eta=0}}
{\left(\frac{d\sigma}{d\Omega}\right)_{\eta=\pi/2}+\left(\frac{d\sigma}{d\Omega}\right)_{\eta=0}}
\,=\,\frac{{\rm sin^2}\theta}{k_0/k_1+k_1/k_0-{\rm sin}^2\theta}
\label{KNequ2}
\end{equation}
where $k_1$ can be computed as function of $k_0$ and $\theta$ with the help 
of the Compton formula:
\begin{equation}
\Delta \lambda = \frac{h}{m_e\, c}\,\left(1-{\rm cos}\theta \right)
\label{cf}
\end{equation}
The left panel of Fig.\ \ref{muFig} shows the modulation factor as function 
of $\theta$ for several initial photon energies. The right panel of the same
figure shows the solid angle and cross section averaged modulation factor as function of energy.
Whereas perpendicular scatterings ($\theta\,=$ $\pi/2$) at low energies 
$E_{\gamma}\ll m_{\rm e}c^2$ have $\mu\,\approx$ 1, the modulation is 
$\ll$1 for scatterings in the forward and in the backward direction.
The modest modulation factors $\mu\,\sim$ 0.5 achieved with the polarimeters 
discussed below largely result from averaging over solid angles. 
Systematic biases are a great concern for X-ray polarimeters. 
An effective means to deal with systematics is a rotation of the 
telescope around the optical axis. The azimuthal distribution of the detected Compton 
events can then be analyzed in celestial coordinates (to extract the 
polarization of the emission) and in detector coordinates (to estimate the 
systematic errors). The RHESSI experiment was rotated at 0.25 Hz, a property 
that was exploited for polarimetry measurements \citep{RHESSI,Cobu:03}.
The method fails for sources with substantial flux variations on the time scale 
of the rotation period (e.g.\ GRBs and fast AGN flares). Another complication are
time variable backgrounds if the background changes on shorter time scales 
than the rotation period.
\subsection{Four Detector Assemblies for Hard X-Ray Polarimetry}
In this section the performance of four hard X-ray polarimeters is compared to each other. 
We discuss the four designs in turn. For descriptions of other hard X-ray polarimeters  
the interested reader might consult \citep{Lei:97,Mizu:05,Mich:08,Vada:10,Bell:10}.
The first design is a detector configuration for the focal 
plane of a Wolter-type (imaging) mirror assembly similar to the one used in 
the {\it HERO} \citep{Rams:02}, {\it HEFT} \citep{Harr:05}, 
{\it InFOC$\mu$S} \citep{Ogas:05} and {\it NuStar} \citep{Harr:05,Harr:10} experiments.
 
The other three polarimeters are large area polarimeters which could be used with a collimator 
for narrow FoV observations of individual sources or with a collimator plus coded 
mask assembly for wide FoV imaging sky surveys. 

We do not evaluate full telescope designs in this paper but limit the discussion to the detector section.
We consider the particular case that the signal strongly dominates over the background and that
the latter is therefore negligible. The main objective of the study is to determine how well the
four designs make use of the signal photons. The design of four full experiments -- including
the optimization of the background shielding and the full simulation of all background sources -- 
is outside the scope of this paper.
%
For all four detector assemblies we assume a photon collection area of 1,600 cm$^2$.
In the case of the narrow FoV polarimeter the 1,600 cm$^2$ would be the effective area
of the mirror assembly (in practice: the effective area of several mirror 
assemblies which focus the X-rays on several identical polarimeters). 
For the wide FoV polarimeters the photon collection area equals the actual area covered with detectors. 
The estimated sensitivities correspond to the
case that the detector assemblies are used with a collimator. If used with a coded mask the
photon collection area is smaller than the detector area as a part ($\sim 50\%$) of the 
detector area is shadowed by the mask elements.
The energy range of the narrow FoV polarimeter is limited at the low-energy end
by the low-energy threshold of the CZT detectors (10 keV), and at the high-energy end by the 
high-energy cut-off of the mirror (80 keV). The energy range of the 
large area experiments is given at the low-energy end by the energy threshold of the
scattering-detectors. The polarimeters could detect photons into the MeV energy range.\\[2ex] 
\begin{figure}[tb]
\begin{center}
\includegraphics[width=4.5cm,angle=270]{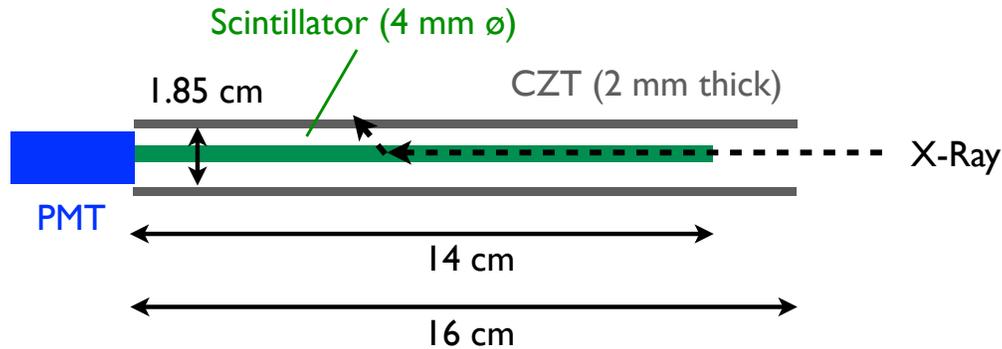}
\caption{\label{D1} Design of the scintillator-CZT polarimeter (Design 1) to be
used in the focal plane of a hard X-ray Wolter-type mirror assembly. The X-rays are focused
on a cylindrical scintillator. The scattered X-rays are detected
with pixelated CZT detectors. The scintillator (energy threshold 2 keV) 
can give a time coincidence signal to tag source events.
}
\end{center}
\end{figure}

The design of the detector assembly for the narrow FoV experiment (Fig.\ \ref{D1}, Design 1) 
uses a cylindrical 14 cm long 4 mm diameter scintillator rod inside a rectangular 16 cm long 
assembly of CZT detectors. Depending on the focal length and diameter of the X-ray 
mirrors, larger diameter rods may be required. The major axis of the scintillator rod 
and the CZT detector assembly are aligned with the optical axis of a Wolter-type mirror assembly. 
The scintillator rod uses a fast scintillator
(EJ-200\footnote{http$://$www.eljentechnology.com$/$datasheets$/$EJ200\%20data\%20sheet.pdf}, 
$\bar{Z}\,=$ 3.4, $\rho\,=$ 1.023 g cm$^{-3}$) with a decay time of 2.1 nsec.
The scintillator rod is read out with a photodetector (PMT or Geiger mode avalanche photodiode) 
at the rear side of the assembly. 
The CZT detector configuration is made of 32 detector units 
each 0.2$\times$2$\times$2 cm$^3$ with a monolithic cathode oriented towards the inside of the 
assembly and 8$\times$8 anode pixels oriented towards the outside. We assume that the scintillator 
and the CZT detectors achieve energy thresholds of 2 keV and 10 keV, respectively. 
Two types of events are recorded: events with one or more triggering CZT pixels (high-background 
events), and events with a trigger of one or more CZT pixels and a trigger of the scintillator 
slab (low-background events). The figures below will show the results obtained for the CZT trigger. 
Additional results for the CZT and scintillator trigger will be described in the text. 
The scintillator/CZT coincidence window should exceed 
$\sim 1\,\mu$s -- the signal formation time in the CZT detectors. 
The interaction depths of the energy depositions in the CZT detectors can be estimated
based on the anode-to-cathode signal ratio \citep{He:96,He:97,Kraw:04}.
The depth information can be used to suppress photons and charged particles that deposit their 
energy close to the outer edges of the CZT detector assembly. The azimuthal
scattering angle is determined from the position of the pixel with the highest 
signal assuming that the photons scatter off the scintillator at the optical axis.
As grazing angle mirror technology is limited to energies $\le$80 keV, we show all results 
for Design 1 over the limited energy range from 10 keV to 80 keV. 
As will be shown below the polarimeter is very sensitive. However, it does not provide
imaging information even though it is located in the focal plane of a Wolter type mirror
assembly. For the description of a less sensitive polarimeter with imaging
capabilities see \citep{Mich:08}.

\begin{figure}[tb]
\begin{center}
\hspace*{-0.25cm}
\includegraphics[width=5.5cm,angle=270]{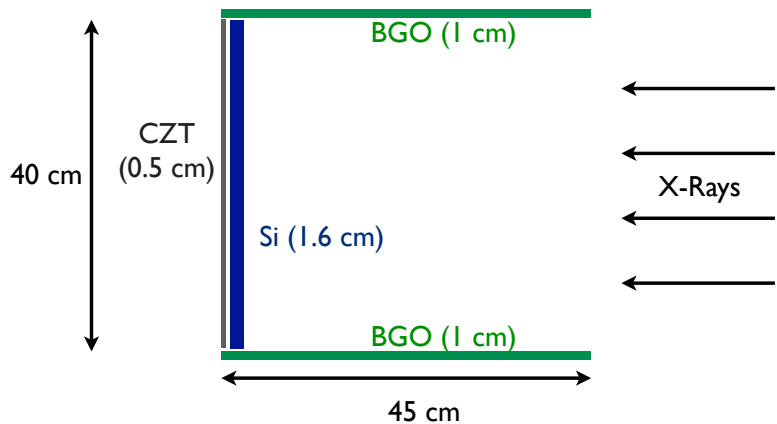}
\vspace*{-0.1cm}
\caption{\label{D2} Design of the Si-CZT-BGO polarimeter (Design 2) to be
used with a collimator or a collimator/coded mask. The assembly includes
stacks of 2 mm thick Si detectors with a combined foot print of 40$\times$40 cm$^2$ 
and a combined thickness of 1.6 cm. Individual Si detector units are 2 mm thick,
have a footprint of $10\times 10$ cm$^2$, and have crossed anode
and cathode strips at a pitch of 2 mm.
The Si detector assembly is followed by an array of CZT detectors 
(footprint: 40$\times$40 cm$^2$, thickness: 0.5 cm, each CZT detector unit: 
0.5$\times$2$\times$2 cm$^3$, 64 anode pixels at a pitch of 2.5 mm, one 
cathode with guard ring). A ring of 1 cm thick, 5.04 cm wide and 45 cm high BGO 
scintillation detectors surrounds the Si and CZT detector assembly.}
\end{center}
\end{figure}

The second detector assembly (Fig. \ref{D2}, Design 2) is similar to the proposed 
polarimeter HX-POL \citep{Kraw:09}. The detectors cover a large area and could be used {\it without} focussing optics either 
with a pencil beam collimator or a coded mask assembly. The design uses 1.6 cm thick stacks 
of Si detectors in front of one layer of 0.5 cm thick CZT detectors.
The Si and CZT detectors are surrounded by a 45 cm high assembly of BGO slabs (each 5.125 cm wide 
and 1 cm thick). The Si detector assembly is made of 4$\times$4 detector stacks.
Each stack is made of eight Si cross-strip detectors. Each Si detector has a 10$\times$10 cm$^2$
footprint and is 2 mm thick. The Si detectors are read out with 
anode and cathode strips at a pitch of 1.4 mm, with the cathode strip direction 
oriented 90 degrees to that of the anode strips. A particle interaction produces a charge 
pulse on the contacts above and below the interaction, giving the x-y position of the event.
The CZT detector array is located 1 cm below the Si detector assembly and is made of 10$\times$10 
CZT detectors, each 5~mm thick and with a footprint of 4$\times$4~cm$^2$. The CZT detectors are 
read out with 16$\times$16 anode pixels at a pitch of 2.5 mm and with four quadratic 2$\times$2 
cm$^2$ cathode segments. 
In the case of HX-POL Application Specific Integrated Circuits (ASICs) are used to read out the 
Si and CZT detectors. Currently, the energy threshold of the Si and CZT detectors are 
12 and 25 keV, respectively \citep{Kraw:09}. Modifications of the ASICs are expected to lead to lower energy 
thresholds of 5 keV and 10 keV, respectively. Each BGO slab would be read out with one 
photomultiplier tube and we assume an energy threshold of 30 keV. 
In the discussion below we distinguish between four types of events: (i) events which trigger one or 
more Si detectors and one or more CZT detectors, (ii) events which trigger $\ge$2 Si detectors,
(iii) events which trigger $\ge$2 CZT pixels (excluding events with hits in
two horizontal or vertical next-neighbor-pixels), and (iv) events which trigger one Si detector or
one CZT pixel and one or more BGO detectors. If a detector component detects two energy depositions
but only one interaction is needed for the reconstruction of the azimuthal scattering angle, we use
the location of the higher energy deposition in the analysis. An optimized analysis which 
uses a ``sequencing algorithm'' to identify the most relevant interactions (e.g.\ \citep{Xu:06}) 
and weighs events according to their information content is outside the scope of this paper.
We note that using sequencing algorithms will only lead to a modest improvement 
of the sensitivity of the polarimeter. The MDP scales roughly proportional to the 
inverse of the square root of the number of properly reconstructed events. 
The fraction $f$ of events which could potentially benefit from a more sophisticated 
analysis (events with more than two interactions) is less than half of all events:
38\%, 28\%, 13\%, and 5\%, for the event types (i)-(iv), respectively. 
The reduction of the MDP achieved with a sequencing algorithm is expected 
to be smaller than by a factor of $\sqrt{1+f}$, as even sophisticated 
sequencing algorithms cannot determine the relative order of the interactions 
with absolute certainty. For example, Xu et al.\ (2006) \citep{Xu:06} discuss 
sequencing algorithms for multiple interactions in CZT detectors. 
For events with two detected interactions, their algorithm identifies 
the first interaction with a probability of 95\% at 100 keV, 80\% at 250 keV, 
and 60\% at 500 keV. For 662 keV primary photons, the algorithm identifies 
the correct order of the interaction for less than half of the events.
Sequencing algorithm are thus expected to lead to a $\ll$20\% reduction of the MDP.
\begin{figure}[tb]
\begin{center}
\hspace*{-0.25cm}
\includegraphics[width=9.6cm,angle=270]{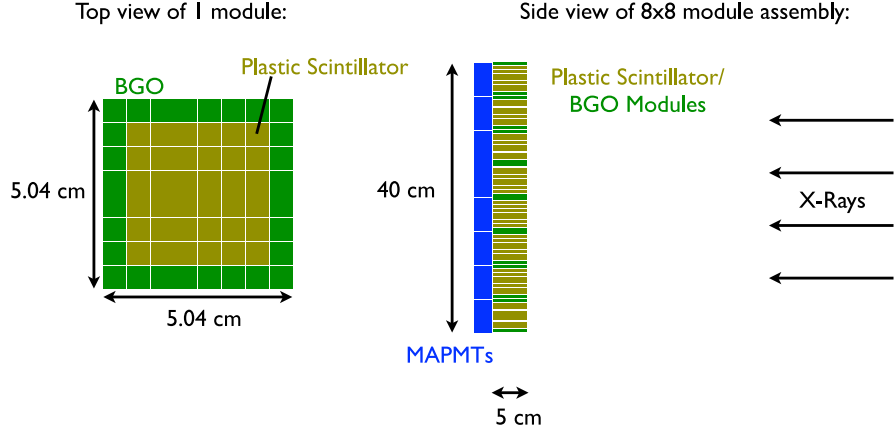}
\vspace*{-3.4cm}
\caption{\label{D3} Design of the plastic scintillator - BGO polarimeter (Design 3) to be
used with a collimator or a collimator/coded mask. The detector assembly is made of 8$\times$8 
polarimeter modules. Each module (left panel) is made of 8$\times$8 scintillator slabs
optically coupled to one 64 channel Multi Anode Photomultiplier (MAPMT).
Each slab has a footprint of 0.5$\times$0.5 cm$^2$ and is 5 cm long.
The inner 36 slabs are made of low-Z plastic scintillator and the outer 28 slabs
are made of high-Z BGO scintillator. The right panel shows how the polarimeters 
modules are exposed to the incoming X-ray radiation.}
\end{center}
\end{figure}

The third detector assembly is similar to the one proposed for the POET and GRAPE experiments 
\citep{McCo:09a,McCo:09b}. 
The detector assembly is made of an 8$\times$8 array of polarimeter modules (Fig.\ \ref{D3}, Design 3). 
Each module consists of an 8$\times$8 array of scintillator slabs (5$\times$5$\times$50 mm$^3$) 
which form a module with 5.08$\times$5.08 cm$^2$ footprint (including gaps between the 
scintillator slabs). Each module is read out by one
Multi-Anode Photomultiplier (MAPMT) with 64 pixels. The 36 central elements of a module are 
low-Z plastic scintillator slabs (EJ-204). These ``scattering'' elements are surrounded by a ``ring'' of 28 high-Z BGO scintillator 
elements. A valid polarimetry event has one hit in the plastic scintillator and one in the 
BGO scintillator. Motivated by the experimental results of \citet{Blos:09} we use energy 
thresholds of 6 keV and 30 keV for the plastic and BGO scintillators, respectively. 
\begin{figure}[tb]
\begin{center}
\hspace*{-0.25cm}
\includegraphics[width=7cm,angle=270]{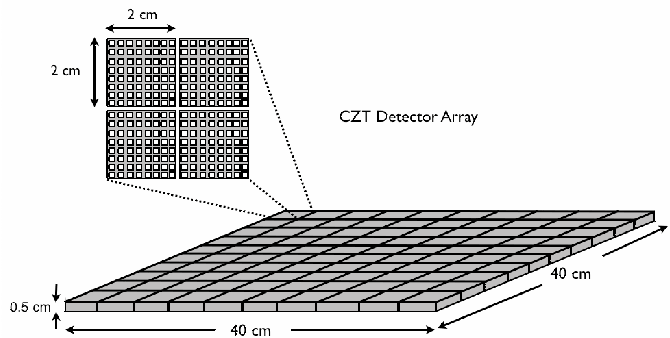}
\vspace*{-1.4cm}
\caption{\label{D4} Design of the CZT polarimeter (Design 4) to be
used with a collimator or a collimator/coded mask. 
The detector assembly is made of an 40$\times$40 cm$^2$ array of 0.5 cm thick 
CZT detectors (each CZT detector unit: 0.5$\times$2$\times$2 cm$^3$,
1024 pixels at 0.6 mm pitch, one cathode with guard ring).}
\end{center}
\end{figure}

The fourth detector assembly is similar to the detector assembly of the Energetic X-ray 
Imaging Space Telescope {\it EXIST~}\footnote{http:$//$exist.gsfc.nasa.gov$/$} \citep{Grin:09} 
-- but smaller. {\it EXIST} is a proposed hard X-ray all sky survey mission with a detector 
area of 4.5 m$^2$ and a tungsten shadow mask for coded mask imaging. We consider here an 
assembly with a detection area of 1,600 cm$^2$ made of 0.5 cm thick CZT detector units 
with a footprint of 2$\times$2 cm$^2$ each (Fig.\ \ref{D4}, Design 4). 
We assume that the CZT detectors are read out with pixels at a pitch of 600 microns 
as planned for {\it EXIST}. The {\it EXIST} design foresees to bump-bond 1024 channel ASICs
directly to the CZT substrates. The CZT detectors are then tiled without gaps 
between adjacent detectors. In the case of {\it EXIST}, the detectors view the 
sky through a hybrid coded imaging mask made of tungsten elements of 0.3 mm 
and 3 mm thickness, and use a partially active and partially passive shield and
collimator assembly.

An assembly of CZT detectors can be used as a Compton polarimeter by analyzing events with
energy depositions in two or more pixels of one or more CZT detector units. One property of 
CZT detectors is that energy depositions below two adjacent pixels will lead
to charge induced on the two pixels (see e.g.\ \citep{Chen:04}). 
Such ``charge sharing events'' can mimic Compton events. We find that the 
exclusion of events with the maximum energy depositions in horizontally or 
vertically adjacent pixels effectively suppresses such charge sharing events \citep{Kraw:08}.
Another concern is the effect of ``weighting potential cross talk'' created by
holes that induce charge on several adjacent pixels when drifting to the cathode 
side of the detector \citep{Kim:09}. Using an ASIC with drift time measurement capability
makes it possible to measure the photon interaction depth, and to properly 
identify true Compton events \cite{Zhan:05a,Zhan:05b,Kim:09}. In practice, a CZT-only polarimeter 
may require such ASICs (presently not foreseen for {\it EXIST}). In our simulations, we 
assume that weighting potential cross talk leads to a negligible number 
of events which mimic Compton events. Following Hong et al.\ \citep{Hong:09}, 
we estimate that a single-pixel energy threshold of 5 keV can be achieved.
The Harvard group (Grindlay et al.) is presently testing first 5 mm thick CZT 
detectors bonded to 1024 channel ASICs. We plan to make a joint study to 
experimentally verify the polarization measurement capabilities of the detectors. 
\subsection{Monte Carlo Simulations and Analysis Methods}
\label{methods}
%
%
We studied the performance of the four detector assemblies based on a Monte Carlo 
study with the GEANT4 package \citep{Agos:03}. We did the simulations for 
Design 1 twice, once with the Standard Electromagnetic Physics List, and, once 
with the Livermore Low-Energy Electromagnetic 
Models\footnote{https:$//$twiki.cern.ch$/$twiki$/$bin$/$view$/$Geant4$/$LoweMigratedLivermore}. 
In the following, we will show the results for Design 1 from the latter package 
as it includes a more detailed modeling of low-energy processes for photon 
energies down to 0.25 keV. We compared the results obtained with both 
simulation packages. The two packages gave slightly different rates of 
detected Compton events and slightly different modulation factors. 
The overall MDPs computed with the two packages were the same up to two 
significant digits. For Designs 2-4 we only did the simulations 
with the Standard Electromagnetic Physics List.
  
For each detector configuration, 2 million 
polarized and 2 million unpolarized photons were simulated.
Photons with energies between 10 keV and 80 keV (Design 1) and 30 keV and 1 MeV (Designs 2-4) 
were generated according to the Crab spectrum measured with the Swift 
Burst Alert Telescope (BAT) telescope
\citep{Tuel:10}:
\begin{equation}
\frac{dN}{dE} \, = \, 10.17 \left( \frac{E}{1\, \rm keV}\right)^{-2.15} \rm ph \, cm^{-2} \, s^{-1} \, keV^{-1}.
\end{equation}
The corresponding $>$10 keV energy flux is 5$\times$10$^{-8}$ erg cm$^{-2}$ s$^{-1}$.
The incident photons and their secondaries are tracked through 
the detector volumes while energy deposits, the interaction locations, and 
interaction processes are recorded. For all simulations the instruments were placed in 
a near-vacuum, similar to a low-Earth orbit environment.
The GEANT4 simulation is followed by a simple detector response simulation.
For each event the code computes the energy deposited in the 
individual detectors and - if applicable - in the pixels, strips 
or slabs of the detectors. A detector signal is used for the analysis
if the deposited energy exceeds the energy threshold of the detector.
If an event triggers more than two detector elements, only the two 
signals with the highest energy depositions are used.  
The events were re-weighted so that the number of detected photons corresponds 
to a 100 ksec observation of the Crab Nebula. The exposure corresponds to a 
$>$10 keV fluence of 5$\times$10$^{-3}$ erg cm$^{-2}$.

We characterize the sensitivity of the polarimeters with the minimum 
detectable polarization (MDP) that can be detected on the 99\% confidence level:
\begin{equation}
{\rm MDP}\,=\,\frac{4.29}{\mu R_{\rm src}}\sqrt{\frac{R_{\rm src}+R_{\rm bg}}{T}} 
\label{mdpEQ}
\end{equation}
where $\mu$ is the modulation factor (compare Equ.\ \ref{modF}), 
$R_{\rm src}$ is the total source counting rate,  $R_{\rm bg}$ is the total 
background counting rate, and $T$ is the integration time. 
The MDP has a value between 0 and 1 and gives the polarized 
fraction of a signal which leads to a detection on the 99\% confidence level.
As mentioned above, the modeling of the background, shielding, and background 
suppression is outside the scope of this paper. We limit our analysis to 
situations where the signal dominates strongly over the background ($R_{\rm src}\gg R_{\rm bg}$) 
and we assume $R_{\rm bg}\,=$ 0 in the following.

For some of the detector configurations discussed in the following the 
$\phi$-distributions show modulations even for unpolarized X-ray beams
because of a non-uniform detector acceptance or because of binning effects (compare Fig.\ \ref{phi}). 
Following \citet{Lei:97} we correct the $\phi$-distributions by normalizing 
the polarized $\phi$-histograms to the unpolarized histograms. 
The number of entries $n_{\rm i}$ in the $i^{\rm th}$ bin of the $\phi$-histogram of 
the polarized signal is scaled according to the equation:
\begin{equation}
n'_{\rm i}\,=\,n_{\rm i}\,\frac{\bar{m}}{m_{\rm i}}
\label{corrEQ}
\end{equation}
where $\bar{m}$ is the mean number of entries per bin of the $\phi$-histogram 
of the unpolarized signal and $m_{\rm i}$ is the number of entries in the $i^{\rm th}$ bin of this histogram.
The prescription flattens the $\phi$-distribution of unpolarized beams and leads to a 
sinusoidal modulation pattern for polarized beams.

We performed a series of simulations to study the effect of the non-uniform detector responses
and binning effects of the Designs 1-4 (see Fig.\ \ref{phi}) on the validity of Equ.\ (\ref{mdpEQ}). 
Non-uniformities can deteriorate the MDP of a polarimeter for given source and background 
detection rates, as certain $\phi$-ranges may be underexposed. The relatively large fluctuations 
in these $\phi$-ranges can lead to relatively large chance polarization degrees. As a consequence,
the minimum polarization degree that can be detected with high statistical significance can increase. 
For each design, we used the simulated $\phi$-histogram for an unpolarized incident photon beam 
as a template to simulate 100,000 statistically independent realizations of the histogram for one
determination of the MDP. Each simulated $\phi$-histogram (bin content and bin error) was corrected 
for the non-uniform detector response according to Equ.\ (\ref{corrEQ}). Taking into account the 
proper error bars, the rescaled histogram was then fitted with the model
\begin{equation}
n(a_0,\phi_0;\phi)\,=\,\bar{n} \left(1+a_0 \, \mu \cos{\left[2(\phi-\phi_0)\right]}\right)
\end{equation}
The value $\bar{n}$ is the mean number of entries in each bin of the histogram: 
$\bar{n}\,=$ $n_0/N$, if $n_0$ events were recorded and the histogram has $N$ bins. 
The values $a_0$ and $\phi_0$ denote the best-fit polarization degree and polarization 
direction, respectively. After histogramming for all 100,000 artificial $\phi$-histograms 
the best-fit $a_0$-values, the MDP was determined as the $a_0$-value being larger 
than 99\% of the histogrammed values. For all four detector designs such simulations were performed for 
the $n_0$-values of 1000, 2000, 5000, 10,000, and 20,000. 
For Designs 1-3 we find an excellent agreement of the MDPs from Equ.\ (\ref{mdpEQ}) with the MDPs 
determined from the simulations within the accuracy of the simulations ($\pm$1\% fractional accuracy).
For Design 4 we find that Equ.\ (\ref{mdpEQ}) underestimates the MDP by a factor of 1.08 -- 
independent of $n_0$. In the discussion below, we thus use Equ.\ (\ref{mdpEQ}) for Designs 1-3. 
For Design 3, we multiply the result of Equ.\ (\ref{mdpEQ}) by the correction factor 1.08.

The performance estimates discussed in the following rely on Monte Carlo simulations 
with the GEANT4 package. The authors of the GEANT4 code tested the 
simulations of Compton processes by comparing simulated results to analytical 
results \citep{Depa:03}. Although the results (i.e.\ the scattering rates and the 
observed modulation factors) are consistent with simple estimates based 
on Eqs.\ (\ref{KNequ1}) and (\ref{KNequ2}), it would be desirable to validate 
the simulations by a comparison to experimental data. A comparative study of 
GEANT4 with the experimentally validated simulation package GEANT3 (including the 
GLECS and GLEPS extensions) \citep{Kipp:04,McCo:04,Lege:05,Blos:06,Blos:09} 
is underway (M.\ McConnell, private communication). 
As described further below, we will compare the simulations of Design 1 to 
experimental data in the near future.
\subsection{Results}
\begin{table}[t]
\begin{tabular}{p{2.8cm}|p{2.6cm}p{2.6cm}p{2.6cm}p{2.6cm}}
\hline
								& Design 1 & Design 2 & Design 3 & Design 4\\ \hline\hline
						
$R_{\rm src}$ [Hz]				& 431.1 & 54.0  & 20.6 & 2.0\\
Peak efficiency 				& 0.85 (70 keV) & 0.3 (145 keV) & 0.11 (145 keV) & 0.035 (416 keV)\\
$\mu$							& 0.52 & 0.38 & 0.41 & 0.61\\
MDP								& 0.13\% & 0.48\% & 0.73\% & 1.7\%\\
\hline \hline
\end{tabular}
\caption{\label{res} Performance of the four detector assemblies:
the rate of Compton events for a Crab-like source $R_{\rm src}$, the peak detection
efficiency and the energy at which this efficiency is achieved, the modulation factor 
$\mu$ and the minimum detectable polarization MDP.    
}
\end{table}
%
%
\begin{figure}[tb]
\begin{center}
\hspace*{-0.5cm}
\begin{minipage}[htb]{6.8cm}
\includegraphics[width=6.8cm]{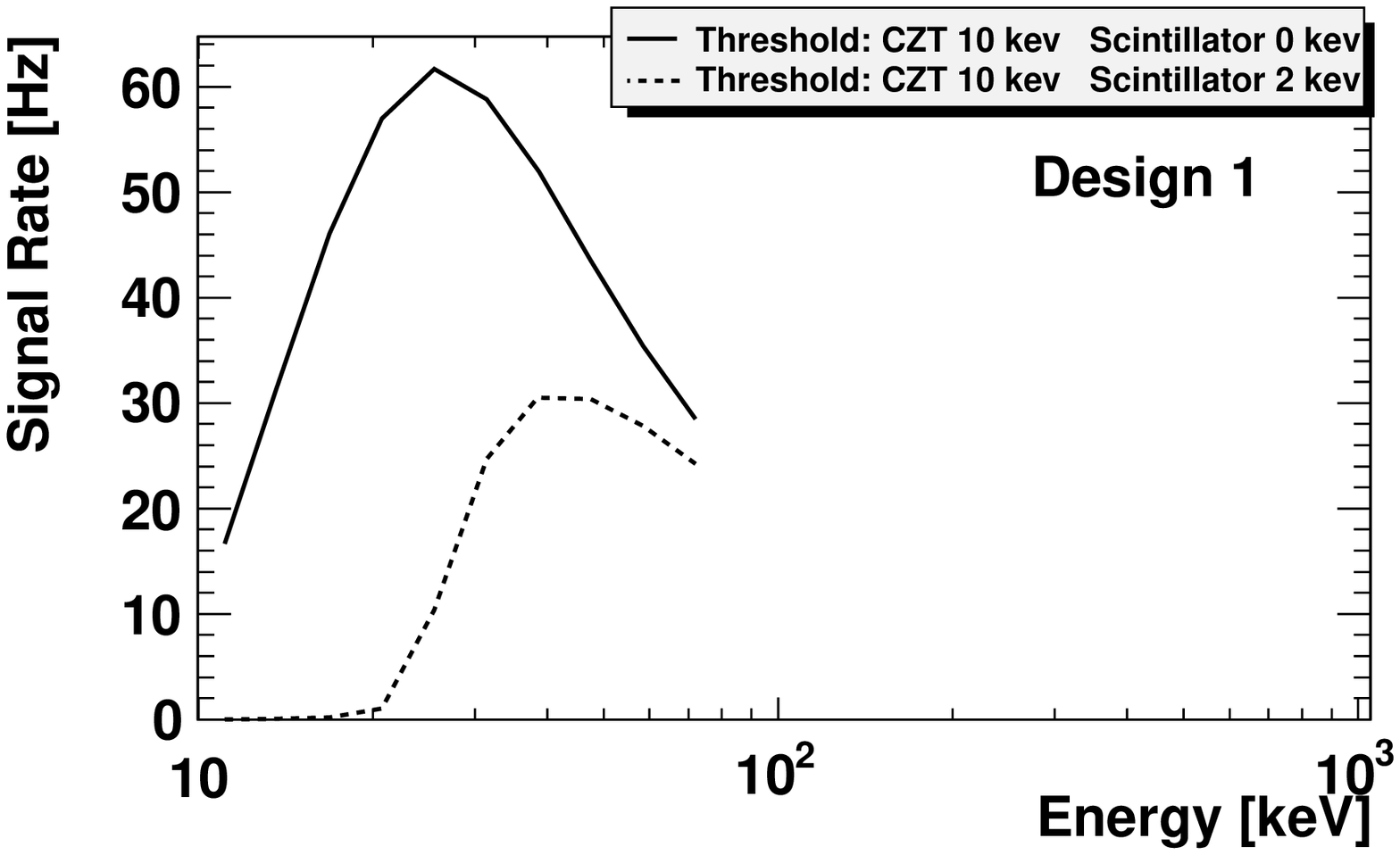}
\end{minipage}
\begin{minipage}[htb]{6.8cm}
\includegraphics[width=6.8cm]{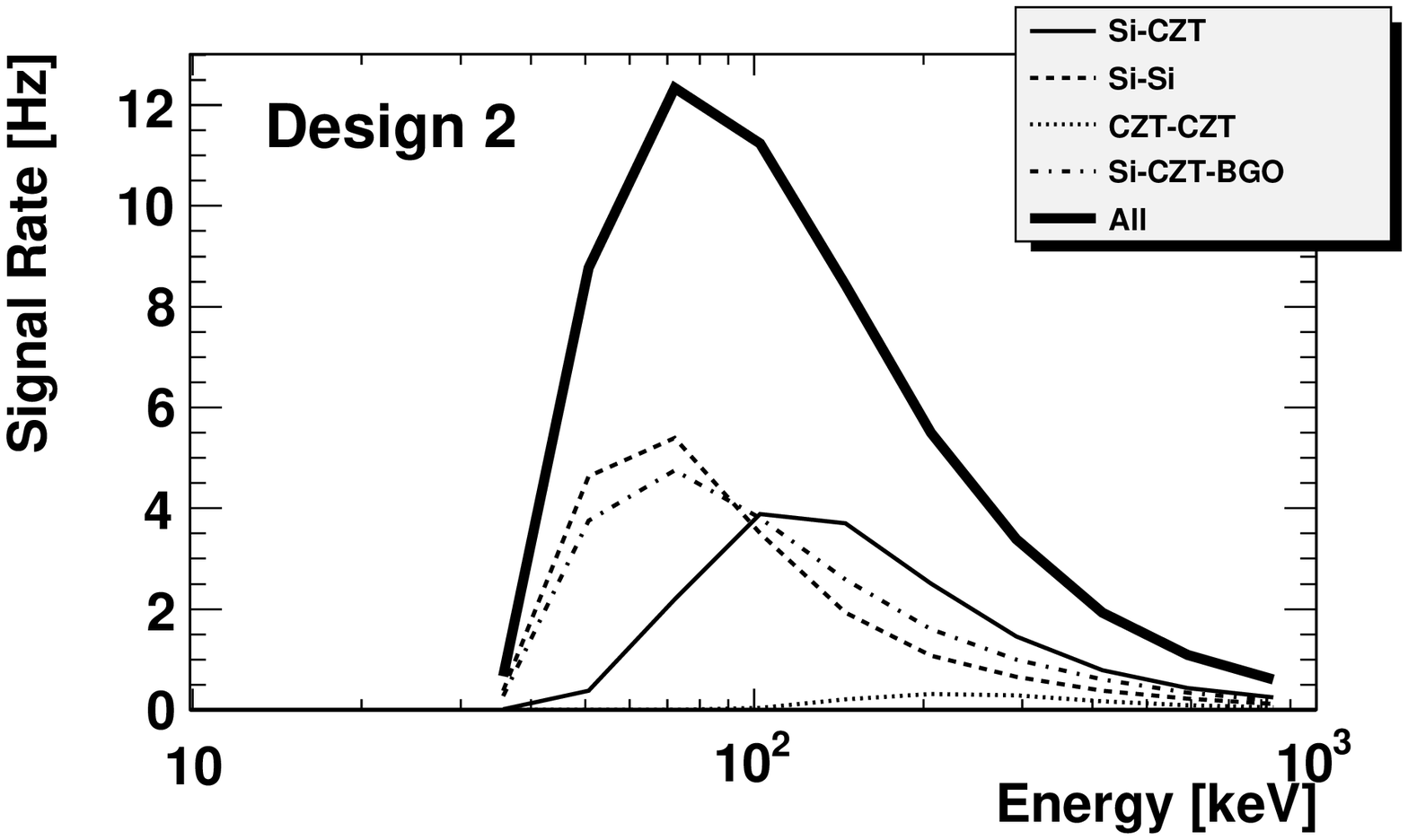}
\end{minipage} 
\hspace*{-0.5cm}
\begin{minipage}[htb]{6.8cm}
\includegraphics[width=6.8cm]{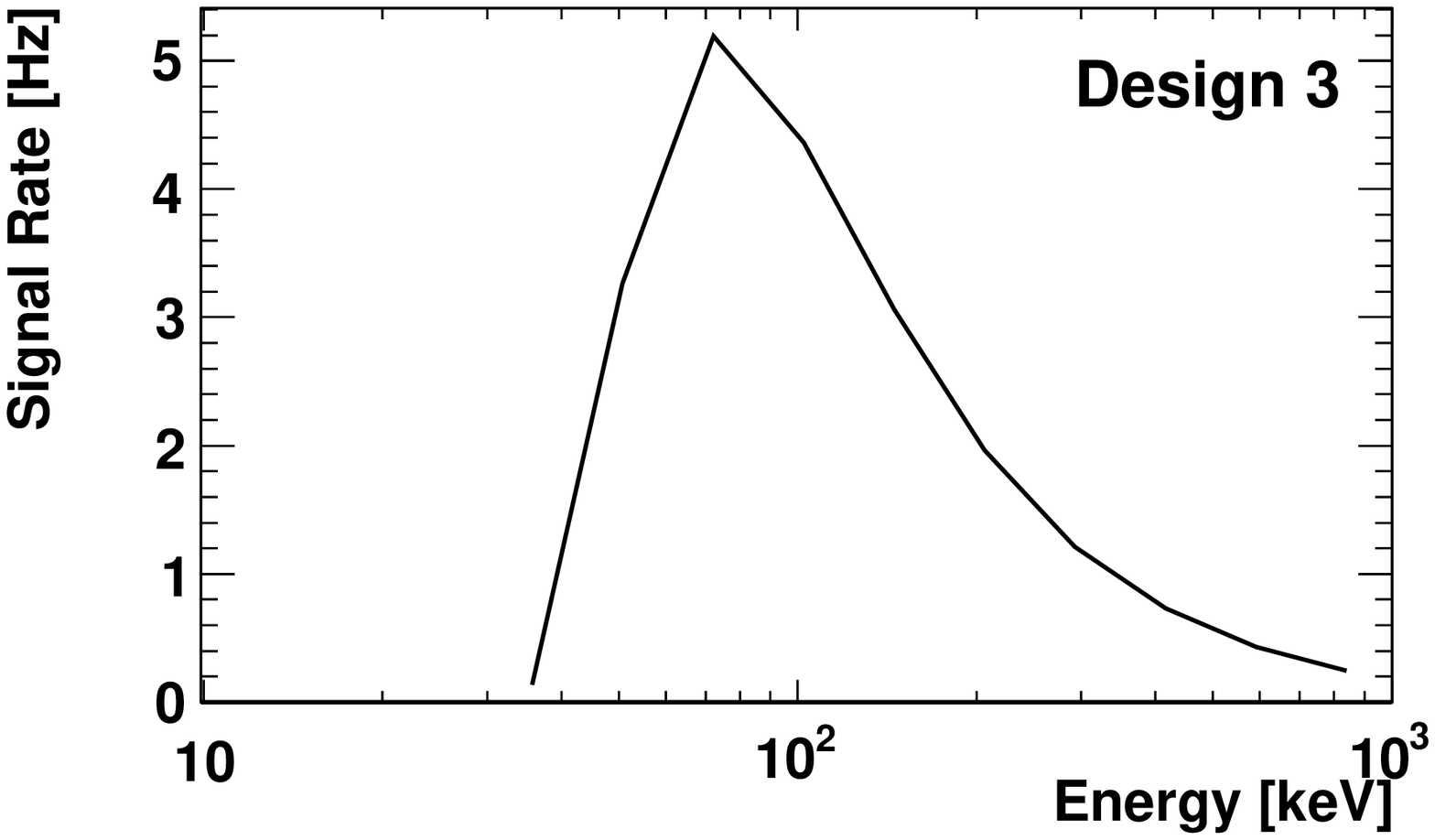}
\end{minipage}
\begin{minipage}[htb]{6.8cm}
\includegraphics[width=6.8cm]{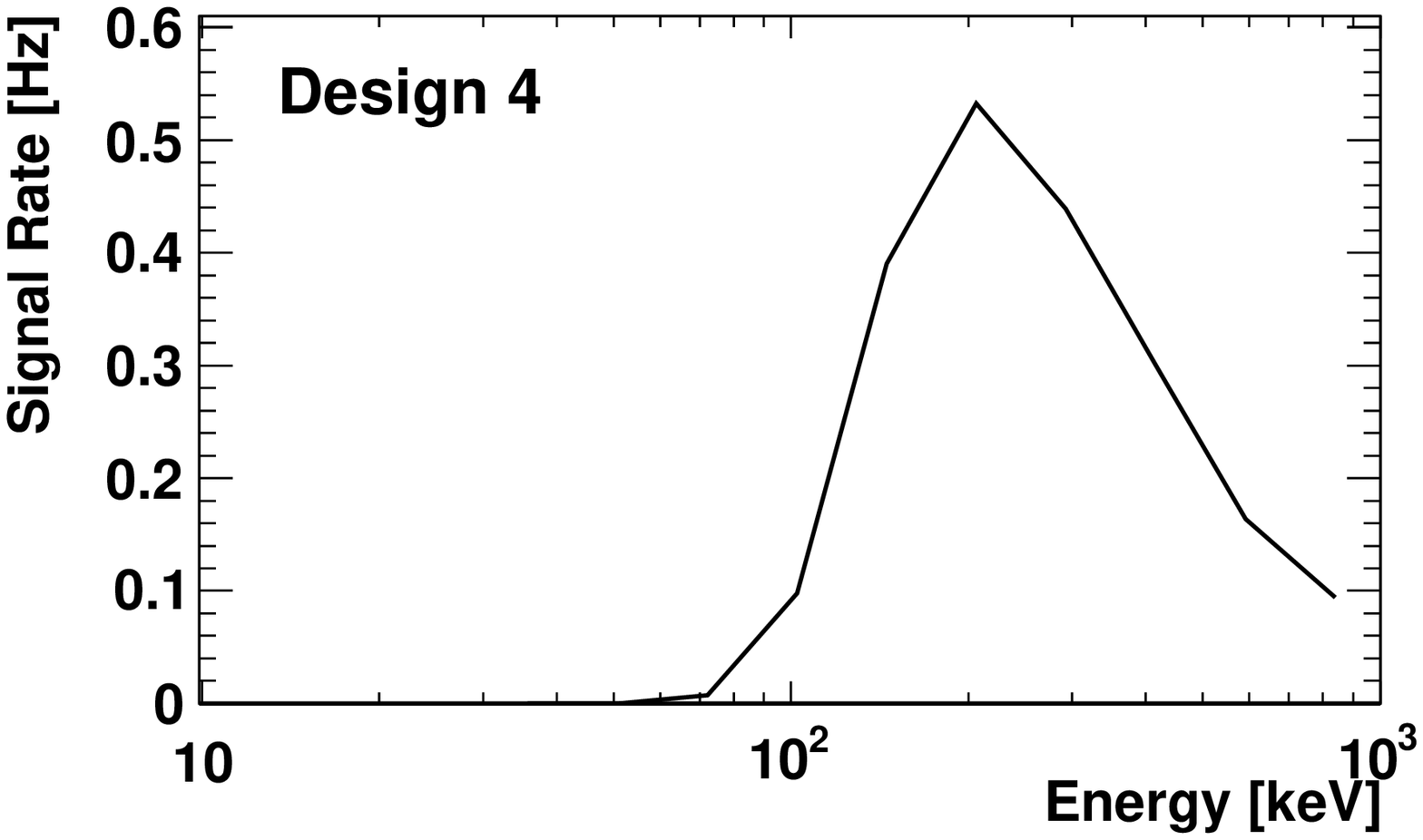}
\end{minipage} 
\caption{\label{detRates} Differential detection 
rates of Compton events for the four detector designs. The figure assumes a 
source with a Crab-like spectrum and flux. For Design 1 the solid line shows the 
rate of events triggering one or more CZT detectors and the dashed line shows
the rate of events triggering one or more CZT detectors and the scintillator.
The latter events will be less contaminated by background.
For Design 2 the solid line shows the total rate of detected 
Compton events; the other lines show the rates for the
different types of events separately.}
\end{center}
\end{figure}
In this section we discuss the performance achieved with the four polarimeters.
The most important results are summarized in Table \ref{res}.
Fig.\ \ref{detRates} shows the detection rates of the four experiments for a 
strong source with a Crab-like flux. Design 1 (the scintillator-CZT assembly 
in the focal plane of focussing X-ray mirrors) achieves the lowest 
energy thresholds. If the only trigger condition is a $>$10 keV hit in one or more CZT 
detectors the assembly is sensitive down to primary photon energies of 10 keV. 
Requiring in addition an energy deposition of $>$2 keV in the scintillator rod 
raises the energy threshold of detected photons to $\sim$20 keV and the peak 
of the differential detection rate increases from 20 keV to 40 keV.
Design 2 (the Si-CZT-BGO detector assembly) has an energy threshold of 
approximately 50 keV. In this case the energy threshold is determined by 
the energy threshold of the Compton scatterer.
The threshold of the Si-Si events and the Si-BGO events is lower than for 
Si-CZT events owing to ``back-scattered'' photons which leave a larger
fraction of their energy in the Si-detectors than the ``forward-scattered'' 
photons that are subsequently photo-absorbed in the CZT detectors.
Whereas Design 3 (the plastic/BGO scintillator detector assembly) 
also achieves a 50 keV energy threshold, Design 4 (the CZT-only experiment)
has a much higher energy threshold of 100 keV owing to the dominance of
photoeffect interactions over Compton scatterings at lower energies.
The integral detection rates for Designs 1-4 are 
431.1, 54.0, 20.6, and 2.0 Hz, respectively (Table \ref{res}).  
 \\[2ex]
%
%
\begin{figure}[tb]
\begin{center}
\hspace*{-0.5cm}
\begin{minipage}[htb]{6.8cm}
\includegraphics[width=6.8cm]{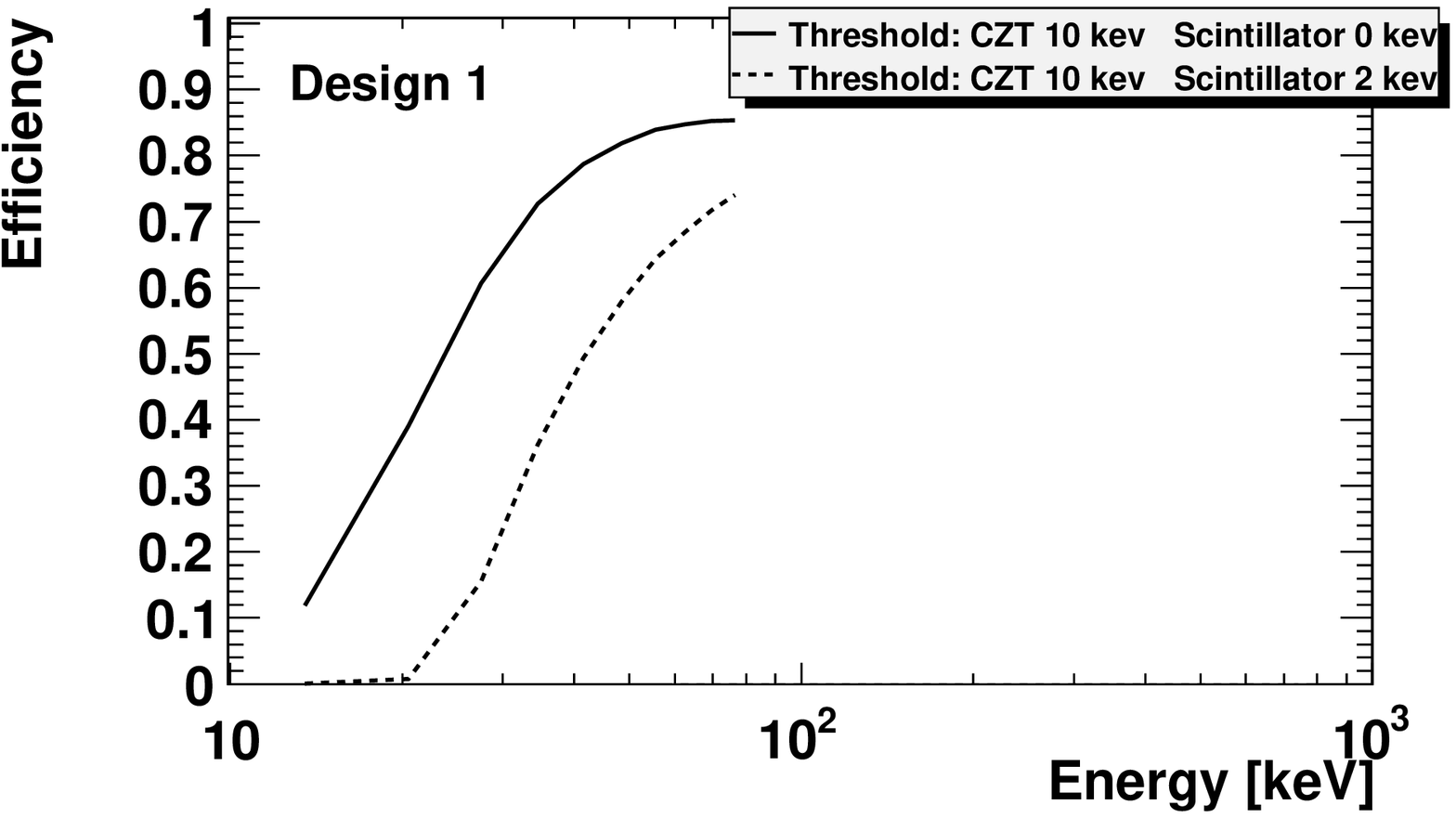}
\end{minipage}
\begin{minipage}[htb]{6.8cm}
\includegraphics[width=6.8cm]{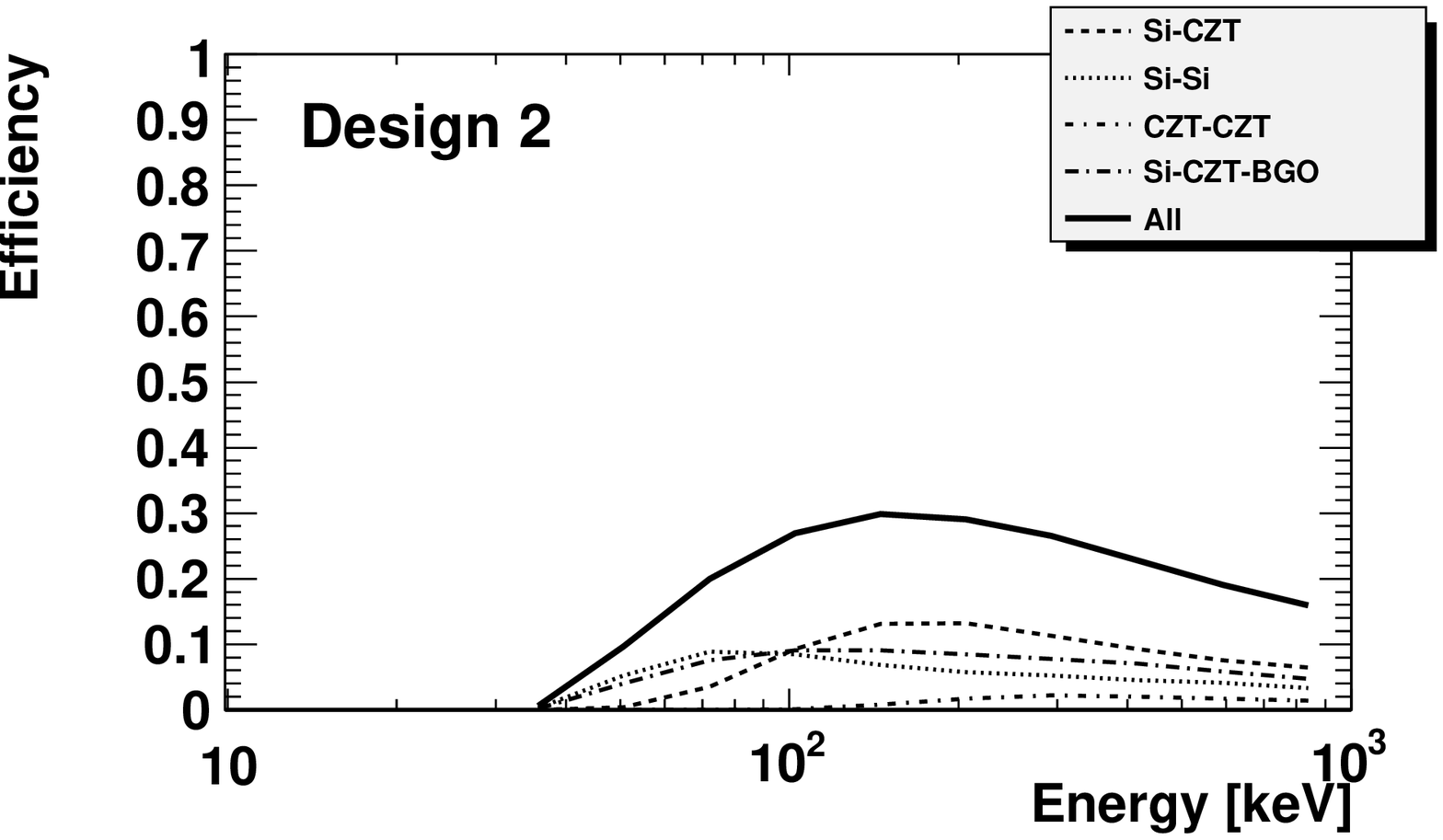}
\end{minipage} 
\hspace*{-0.5cm}
\begin{minipage}[htb]{6.8cm}
\includegraphics[width=6.8cm]{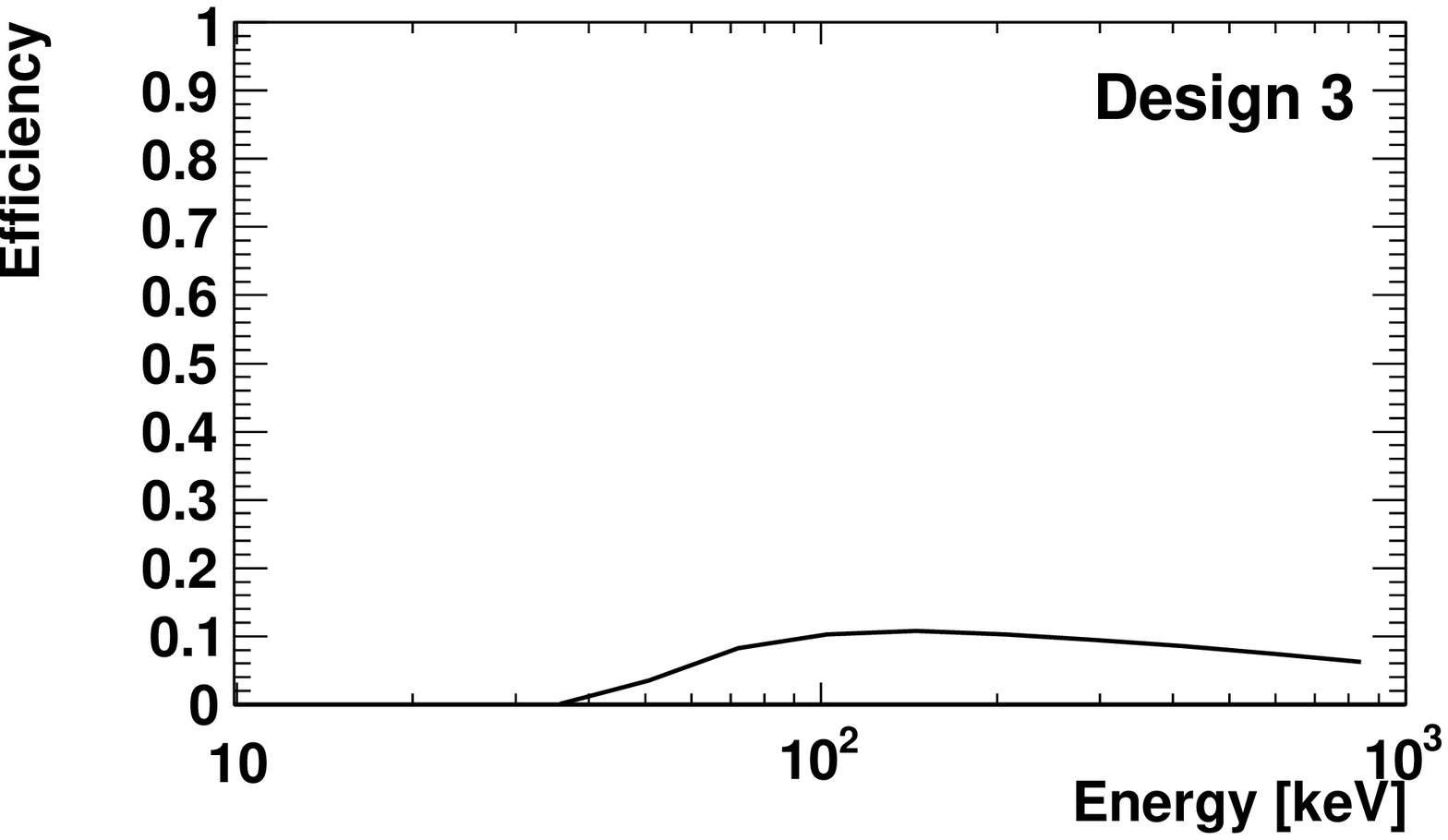}
\end{minipage}
\begin{minipage}[htb]{6.8cm}
\includegraphics[width=6.8cm]{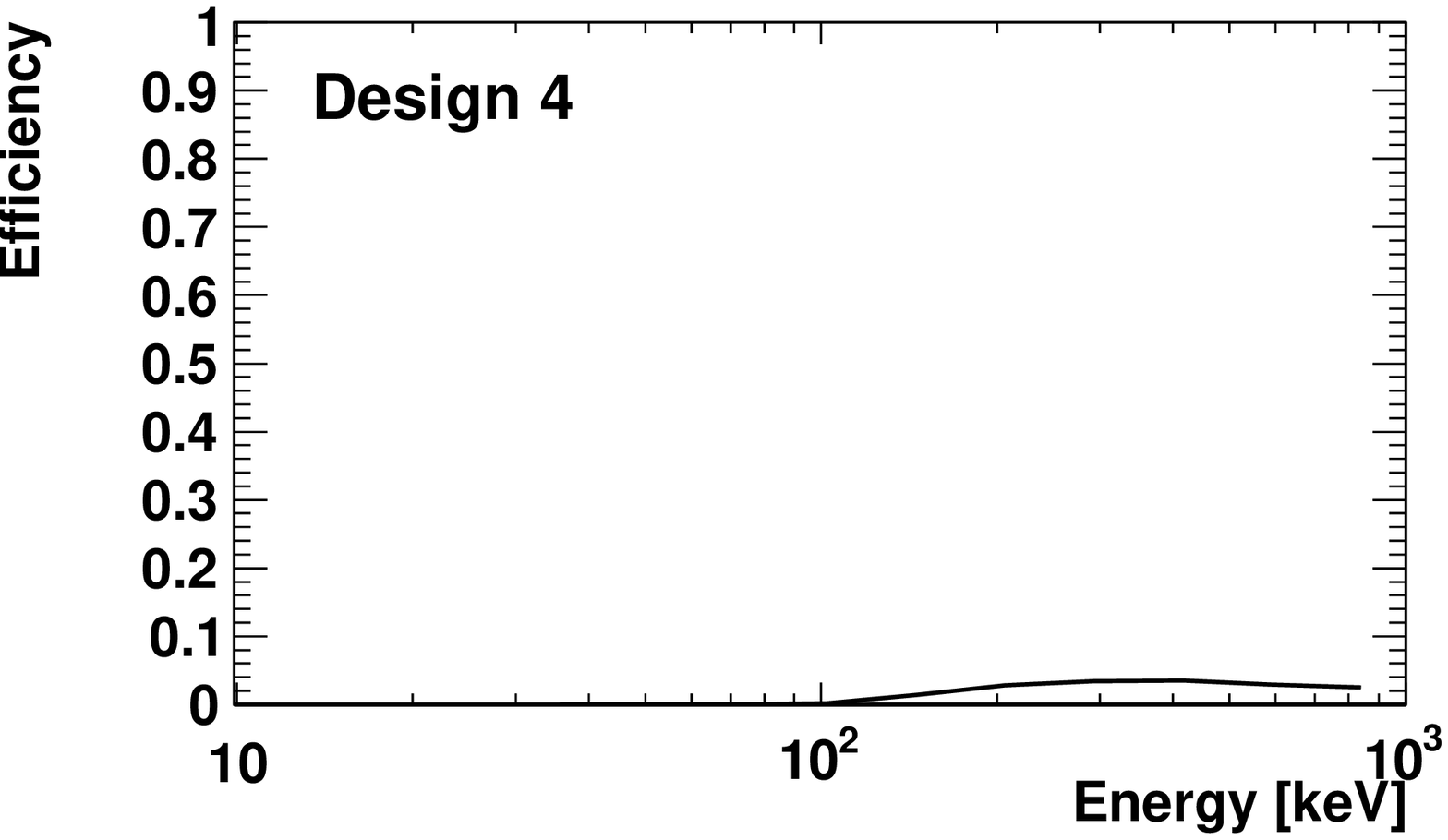}
\end{minipage} 
\caption{\label{eff} Differential efficiencies for the 
four detector assemblies. The efficiency is defined here as the 
fraction of all photons hitting the detector assembly which
causes a trigger and enters the polarization analysis.
The different lines show the results for different types of events 
as in Fig.\ \ref{detRates}.}
\end{center}
\end{figure}
The detection efficiencies of the four detector configurations are shown in 
Fig.\ \ref{eff} and the peak efficiencies are listed in Table \ref{res}. 
The efficiency is defined here as the fraction of photons 
impinging on a detector assembly that triggers the instrument and enters the 
polarization analysis. Design 1 achieves high peak detection efficiencies of 
85\% at 70 keV. All source photons hit the scintillator rod at approximately the 
same location. Owing to the length of the rod a large fraction of the 
photons Compton scatter in the low-Z material. The small-diameter of the 
rod makes it possible that a large fraction of the scattered photons escape. 
Finally, the large solid angle coverage of the high-Z CZT detector assembly
assures that a large fraction of scattered photons is photo-absorbed. 
The net detection efficiency of Design 2 is lower than for Design 1 and
reaches a peak-value close to 30\% at 145 keV. The efficiency peaks at
higher energies owing to the higher Z and the higher trigger threshold 
of the Si Compton scatterer. The peak efficiency results from the conspiracy 
of several facts including the partial transparency of the Si detectors
and the CZT detectors at $>$100 keV energies and a rather high probability 
that photons backscatter from the Si and escape absorption in the CZT and BGO.
The peak detection efficiency of Design 3 is about 11\% at 145 keV. In this detector configuration
events are lost owing to the facts that (i) the low-Z plastic scintillator covers only
56\% of the area and 44\% of the events hit the high-Z BGO scintillator first, 
(ii) Compton scattered photons may be absorbed in the low-Z plastic scintillator
before reaching the high-Z BGO scintillator, (iii) events can escape 
towards the front and rear sides of the assembly. 
The detection efficiency of Design 4 maxes out at relatively low values of 3.5\% at 416 keV.
At the low-energy end the low efficiencies result from the dominance of photoeffect
interactions. Even if a photon Compton scatters it is quite likely that it is absorbed
before it can propagate to the next admitted pixel. At higher energies the CZT detectors 
become partially transparent, and a large fraction of photons escapes the detectors 
towards the front or the rear of the assembly.\\[2ex]
%
%
\begin{figure}[tb]
\begin{center}
\hspace*{-0.5cm}
\begin{minipage}[htb]{6.8cm}
\includegraphics[width=6.8cm]{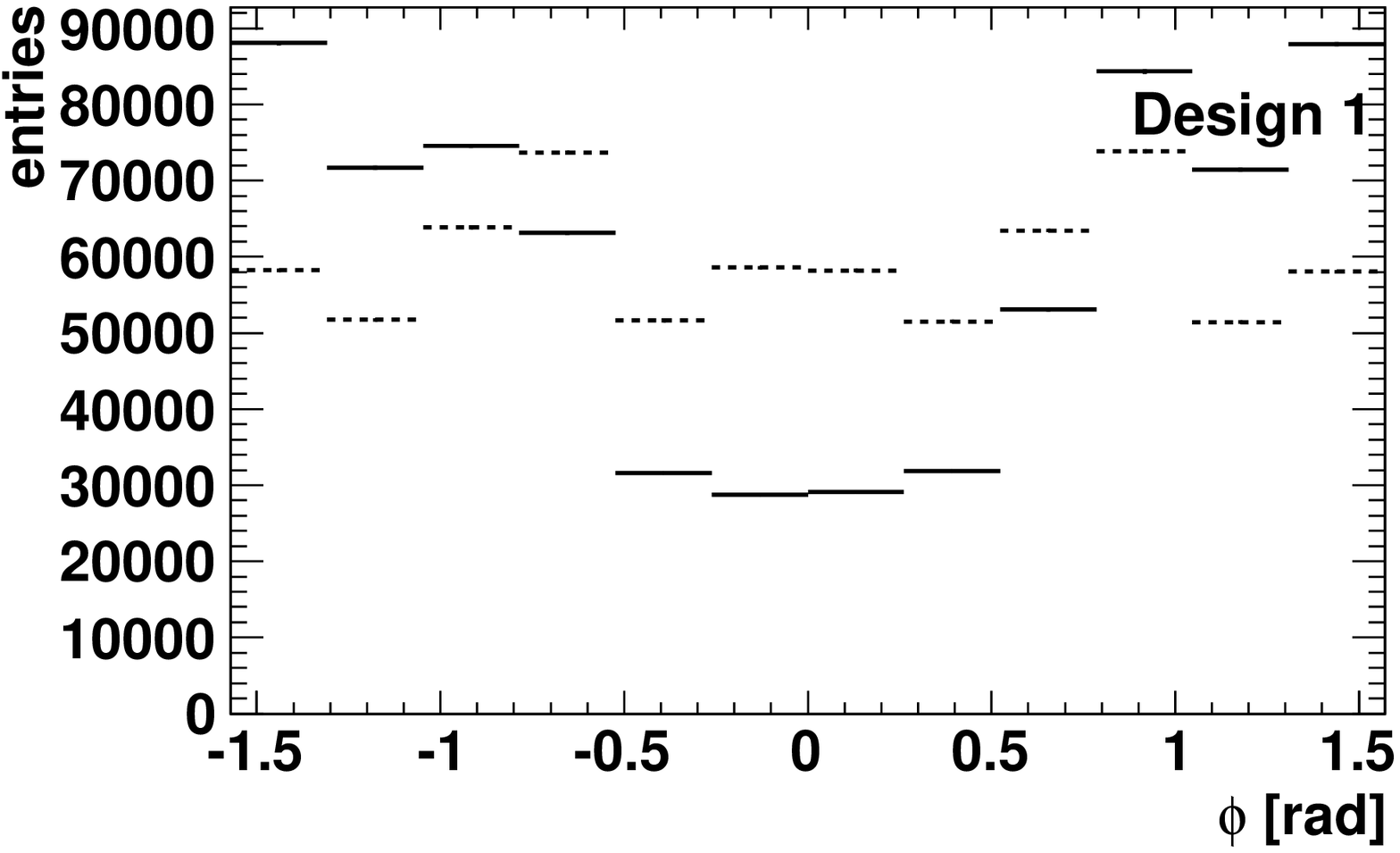}
\end{minipage}
\begin{minipage}[htb]{6.8cm}
\includegraphics[width=6.8cm]{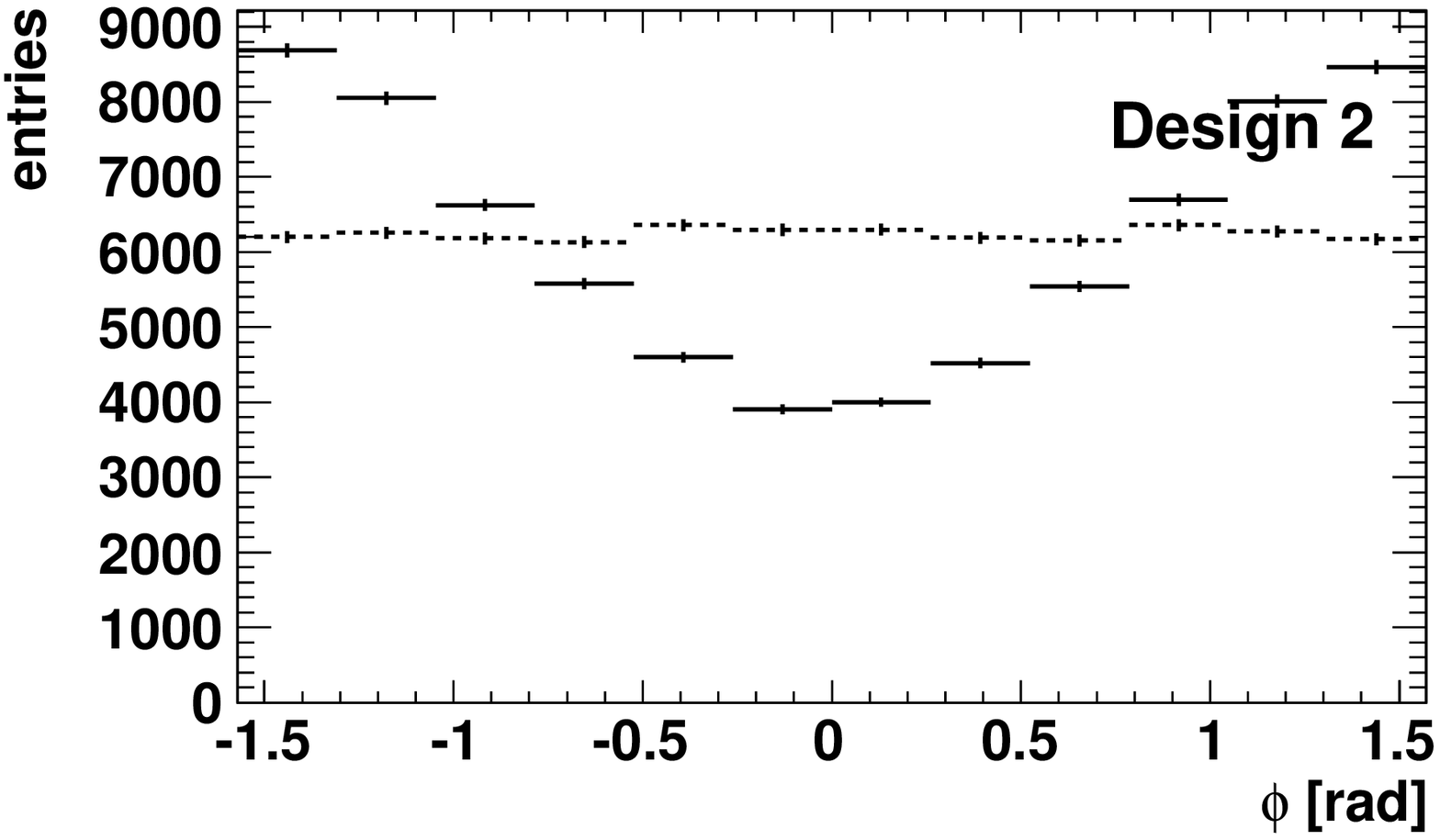}
\end{minipage} 
\hspace*{-0.5cm}
\begin{minipage}[htb]{6.8cm}
\includegraphics[width=6.8cm]{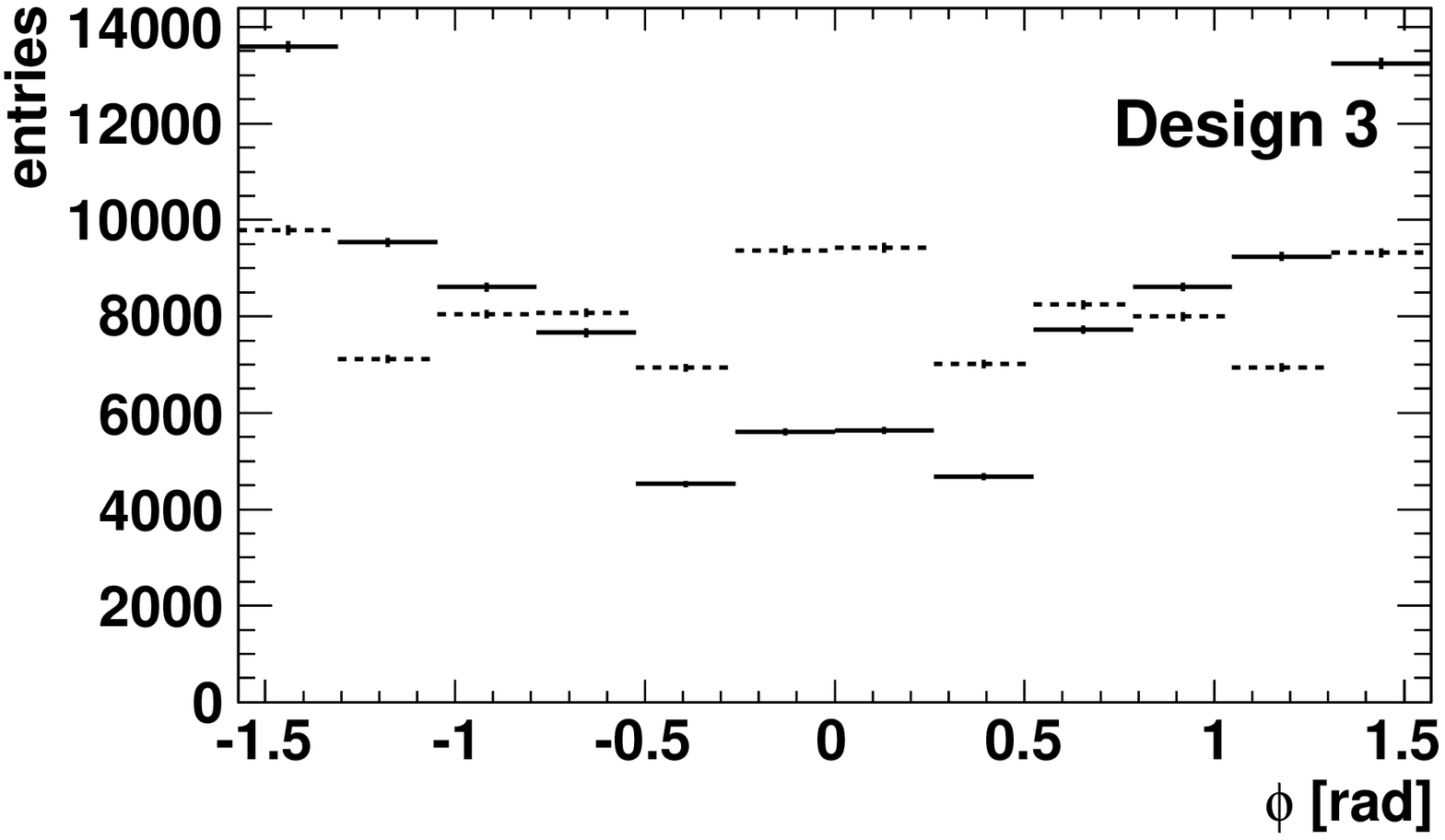}
\end{minipage}
\begin{minipage}[htb]{6.8cm}
\includegraphics[width=6.8cm]{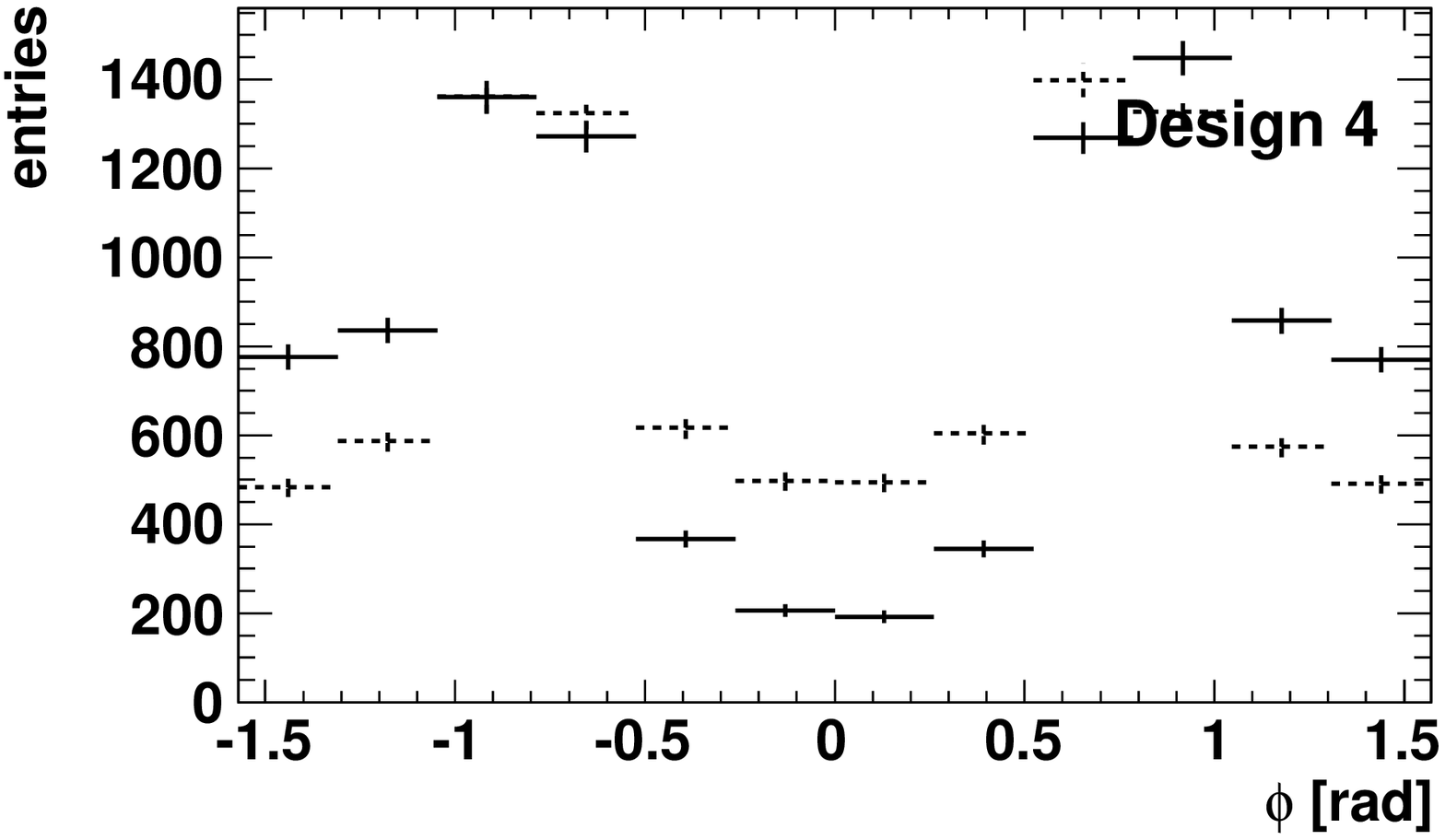}
\end{minipage} 
\caption{\label{phi} Distribution of azimuthal scattering angles for the four detector
assemblies for polarized (solid lines) and unpolarized (dashed lines) incident beams.
For Design 1 we used events which triggered one or more CZT detectors. 
For Design 2 the distribution is shown for events with one interaction in a Si 
detector and one or more interactions in a CZT detector.
}
\end{center}
\end{figure}
Figure \ref{phi} shows exemplary azimuthal scattering distributions for the four detector
configurations for unpolarized and polarized X-ray beams (before correction for non-uniformities). 
Here and below the graphs of Design 1 show the results for events triggering one or more
CZT detectors (no scintillator trigger required); for Design 2 the results are shown for all event types.

The $\phi$-distribution for an unpolarized beam shows some modulation (17\%) for Design 1 
owing to the large pixel size (2.5 mm) and associated binning effects. We 
confirmed that the binning effects go away for finer pixelated detectors 
by simulating Design 1 also with CZT detectors with pixel pitches of 
350 microns and 600 microns. The analysis of the simulation results confirmed  
that the achieved MDP does not depend on the pixel pitch. For Design 2 the 
unpolarized $\phi$-distribution for the Si-CZT events (upper right panel) 
is rather uniform. The unpolarized 
$\phi$-distributions for the Si-Si events, Si-BGO events and CZT-BGO
events (not shown) are also flat. An exception is the unpolarized $\phi$-distribution of the
CZT-CZT events (not shown, see the panel of Design 4 for a similarly modulated 
$\phi$-distribution). 
The pronounced modulation (16\%) of the unpolarized $\phi$-distribution of Design 3
(lower left panel) results from the geometry of the detector configuration, 
i.e.\ the fact that events scattered in the plastic scintillators encounter 
low-Z and high-Z material with effective thicknesses that depend on the azimuthal scattering angle. 
A random distribution of low-Z and high-Z scattering elements in the detector plane may remove these non-uniformities. 
The rather large modulation (40\%) of the unpolarized $\phi$-histogram of Design 4 results from the limited
propagation length of scattered photons in the CZT material. The histogram clearly shows
event accumulations at $\phi\,=$ $\pm\frac{1}{4}\pi$ corresponding to hits
in diagonal next-neighbor pixels.

%
%
\begin{figure}[tb]
\begin{center}
\hspace*{-0.5cm}
\begin{minipage}[htb]{6.8cm}
\includegraphics[width=6.8cm]{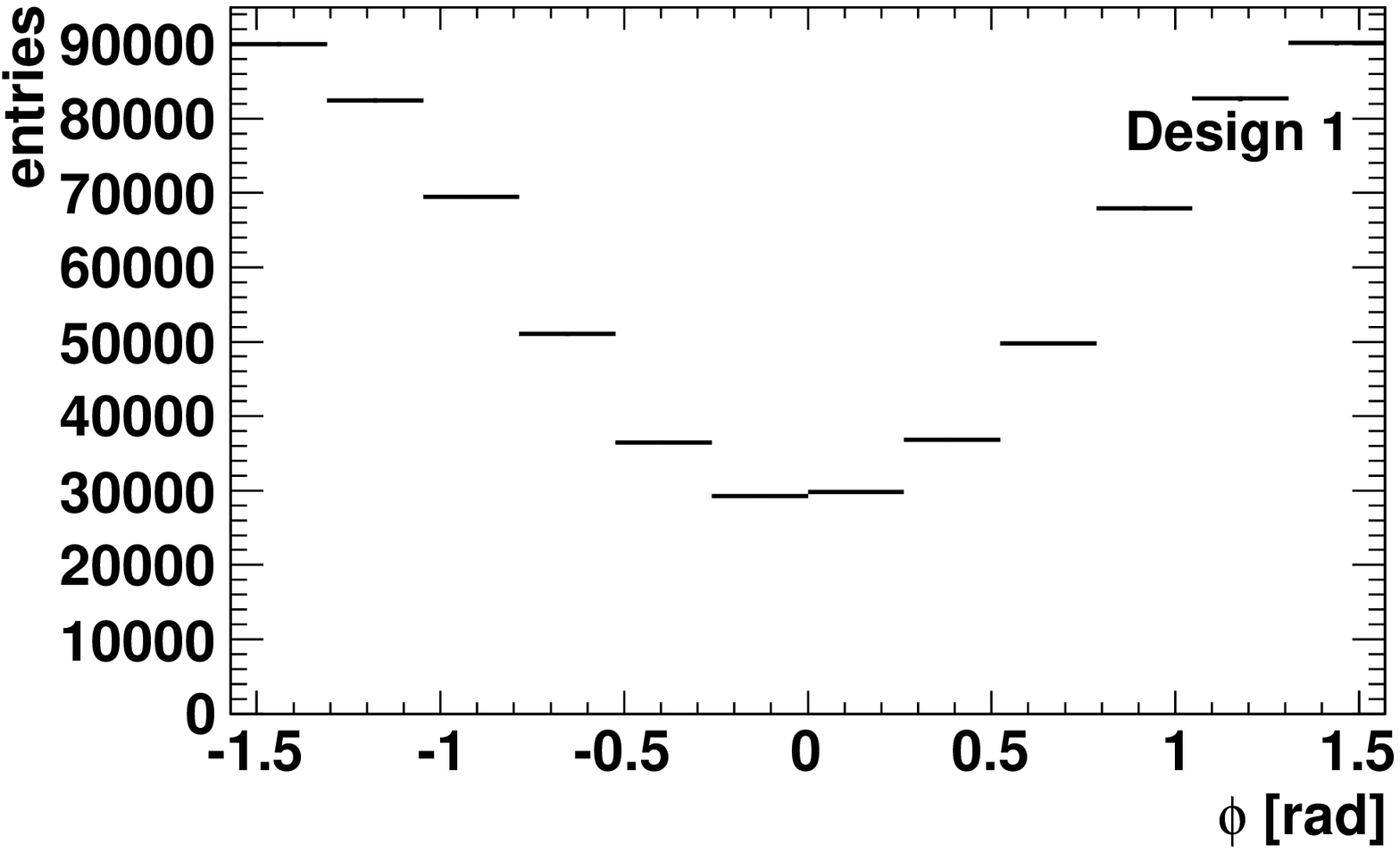}
\end{minipage}
\begin{minipage}[htb]{6.8cm}
\includegraphics[width=6.8cm]{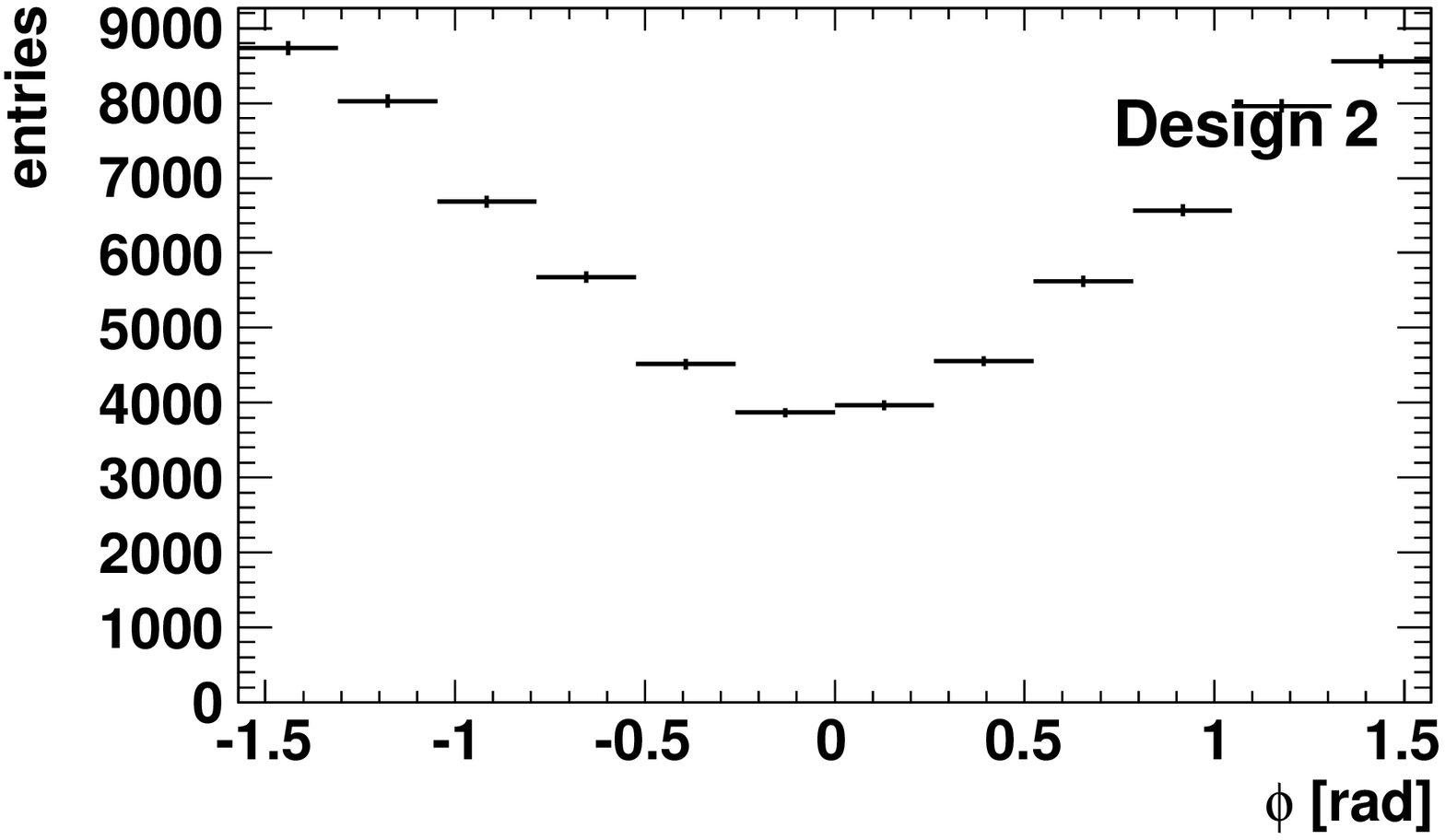}
\end{minipage} 
\hspace*{-0.5cm}
\begin{minipage}[htb]{6.8cm}
\includegraphics[width=6.8cm]{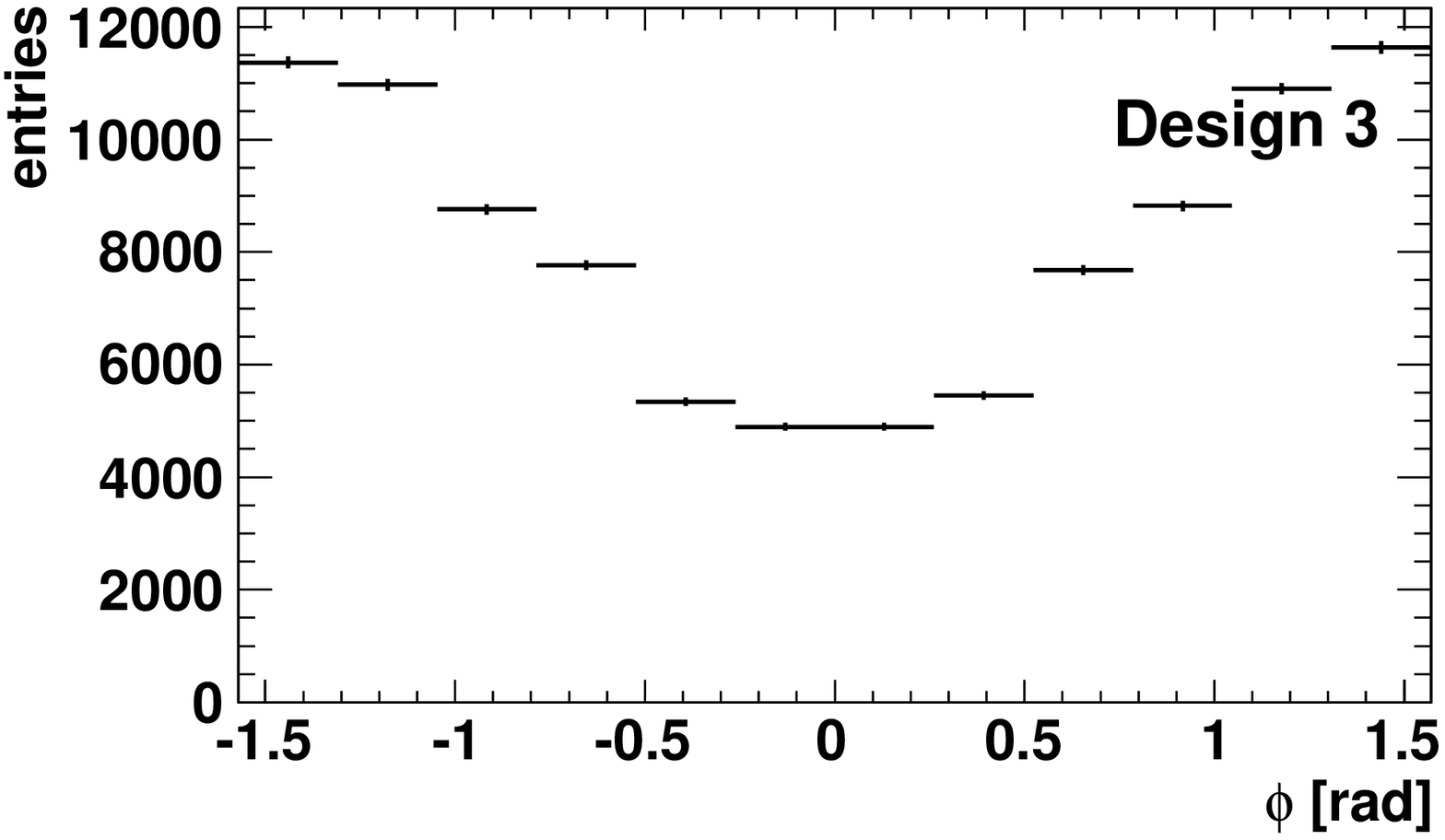}
\end{minipage}
\begin{minipage}[htb]{6.8cm}
\includegraphics[width=6.8cm]{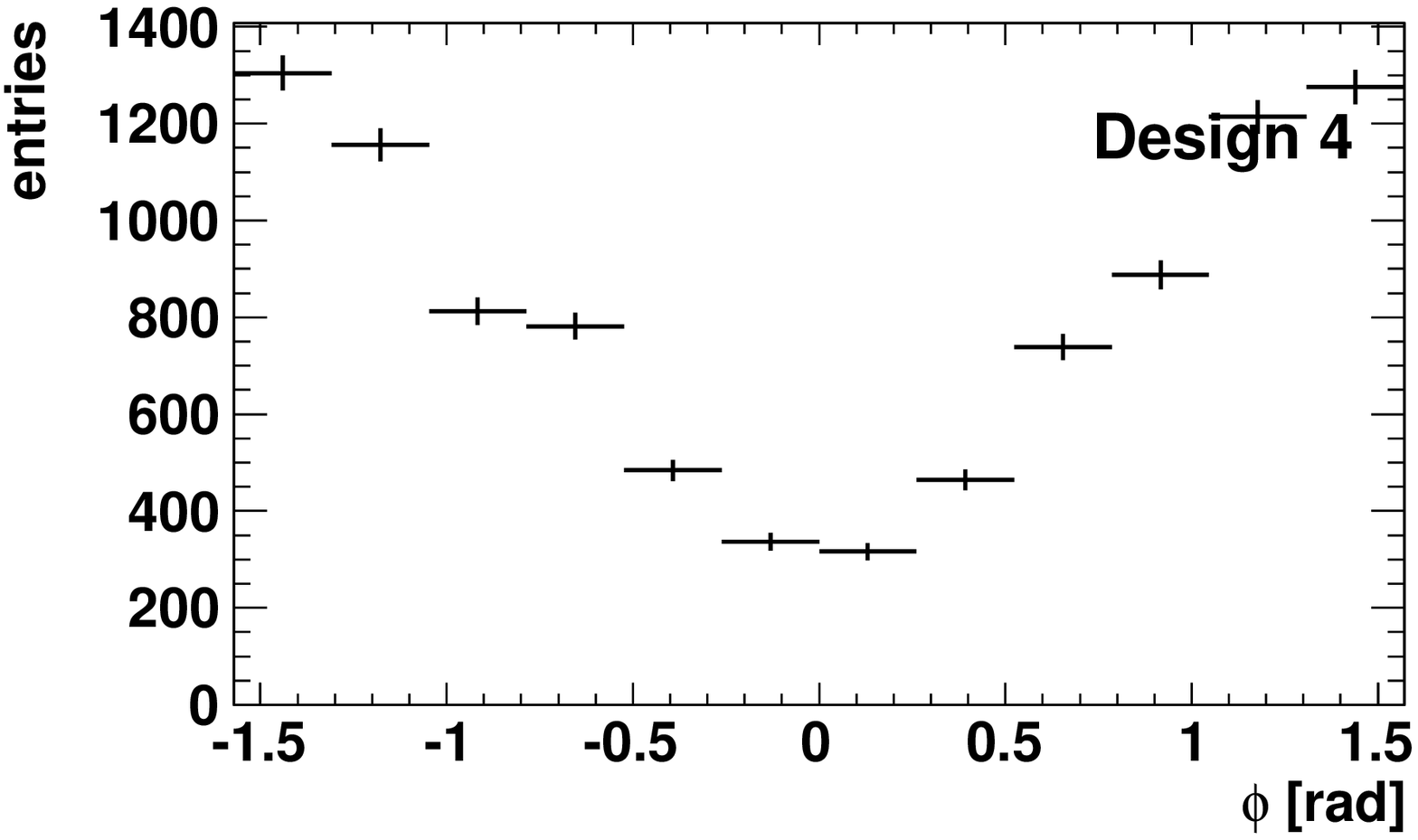}
\end{minipage} 
\caption{\label{phiC} The distributions of azimuthal scattering angles 
for a polarized X-ray beam after correcting for instrumental non-uniformities
for the four detector assemblies. The distributions for unpolarized beams
are not shown as they are constant owing to the correction.  
}
\end{center}
\end{figure}
The correction discussed in Section \ref{methods} flattens the $\phi$-distributions of the
unpolarized beams and leads to a sinusoidal modulation of the $\phi$-distributions of the
polarized beams (Fig.\ \ref{phiC}). After the correction, the modulation factors 
are typically $\sim$0.5 (see Table \ref{res}). 
Detector configurations for which the polar scattering angles close to $\pi/2$ are more common
(e.g.\ in the case of the CZT-only polarimeter) achieve higher $\mu$-values.
\begin{figure}[tb]
\begin{center}
\hspace*{-0.5cm}
\begin{minipage}[htb]{6.8cm}
\includegraphics[width=6.8cm]{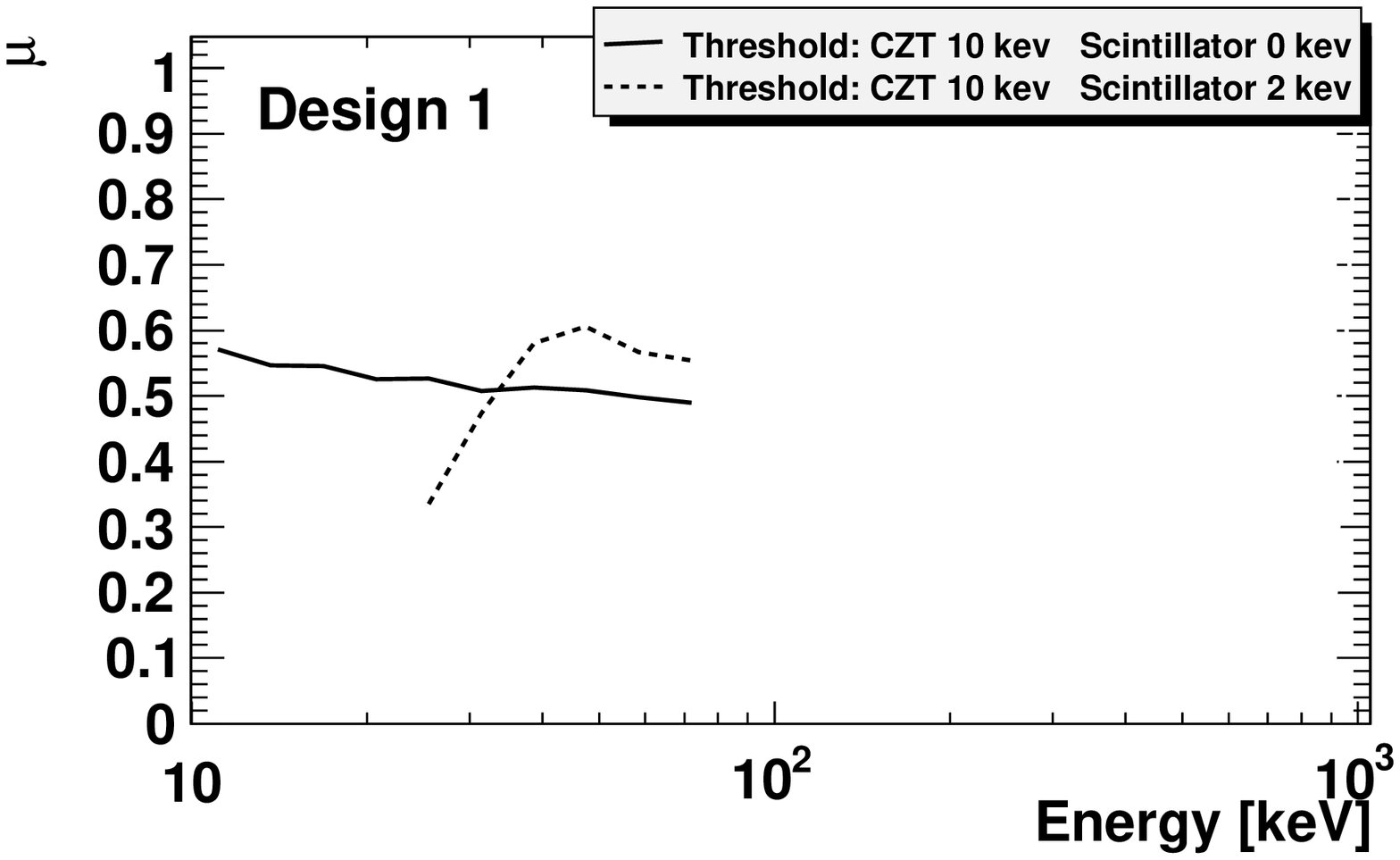}
\end{minipage}
\begin{minipage}[htb]{6.8cm}
\includegraphics[width=6.8cm]{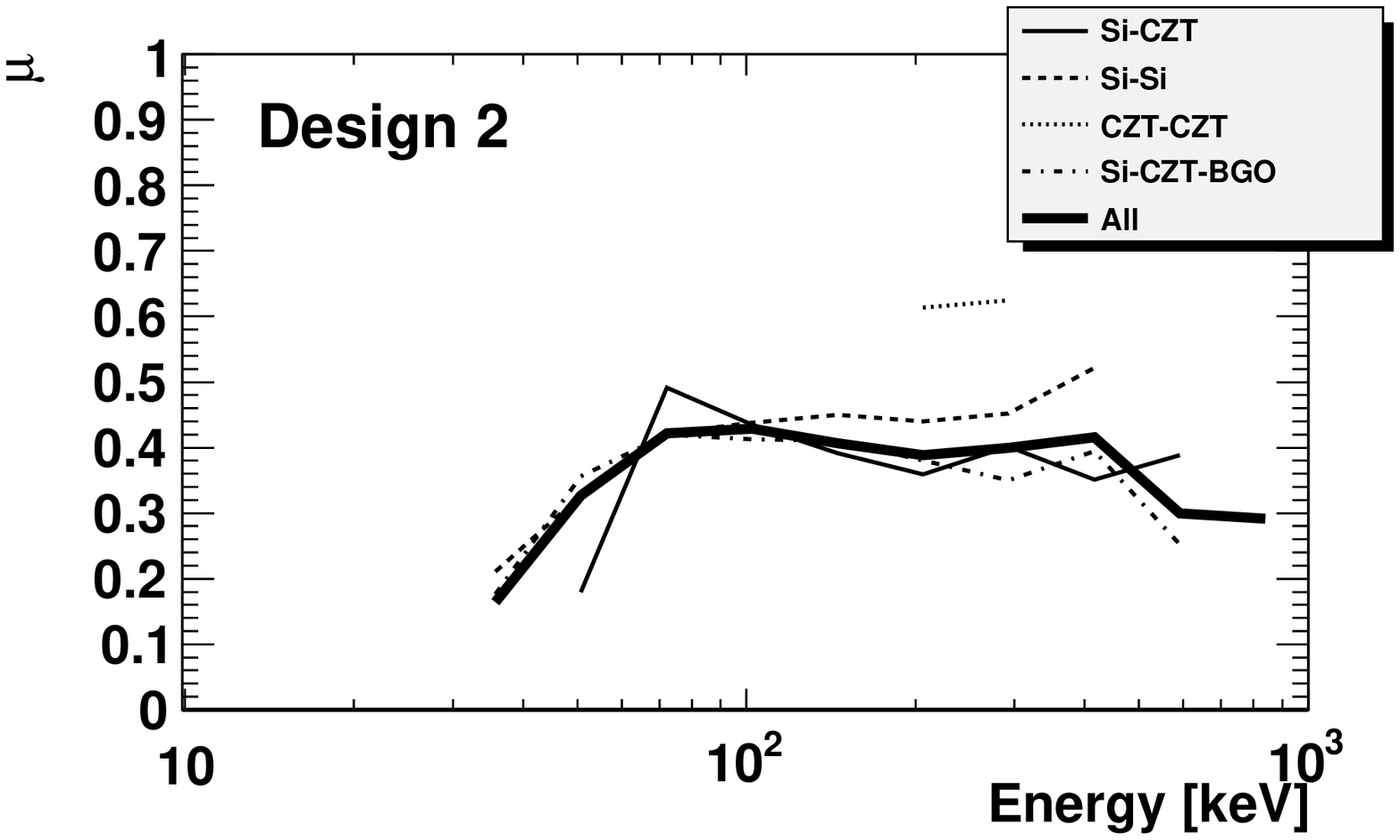}
\end{minipage} 
\hspace*{-0.5cm}
\begin{minipage}[htb]{6.8cm}
\includegraphics[width=6.8cm]{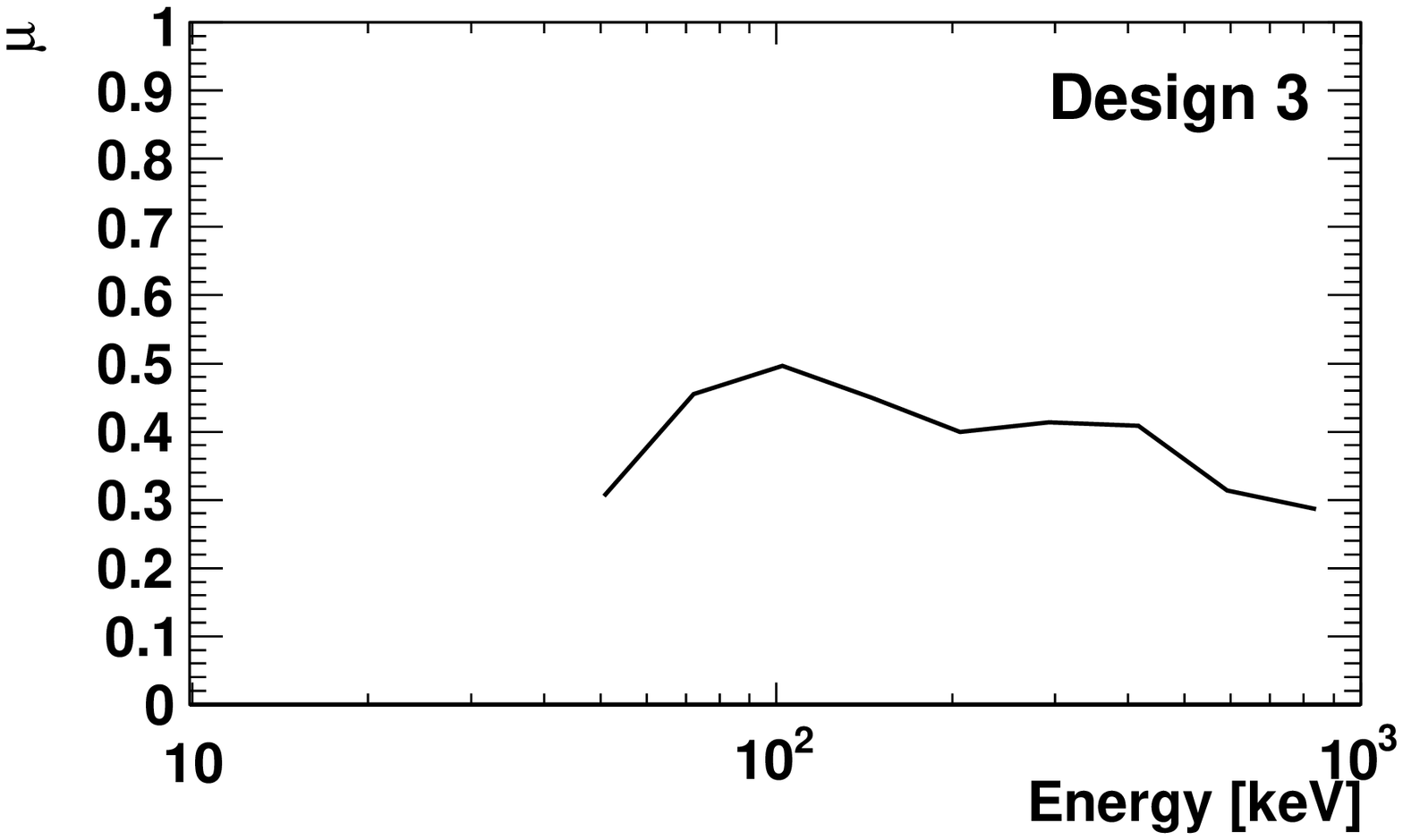}
\end{minipage}
\begin{minipage}[htb]{6.8cm}
\includegraphics[width=6.8cm]{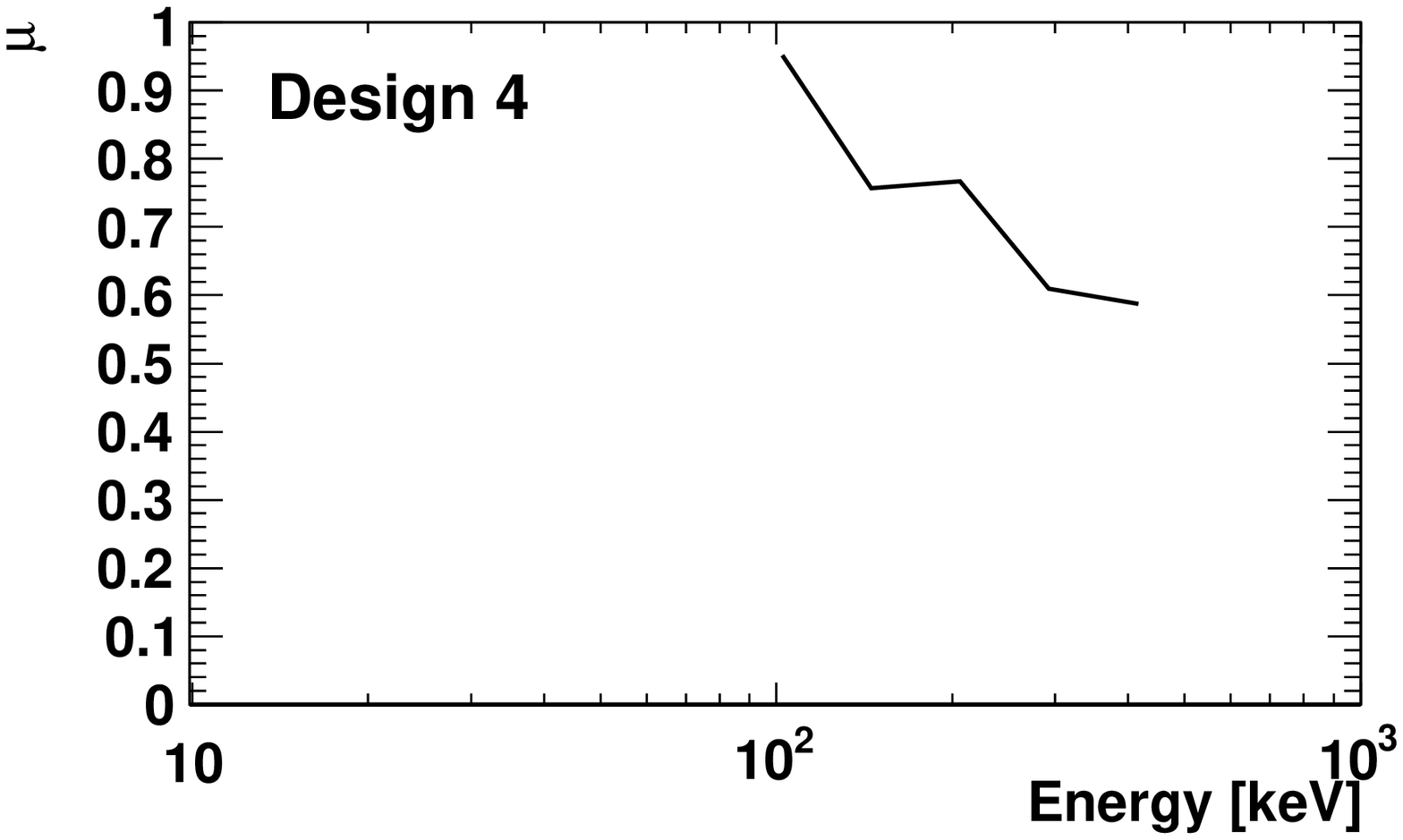}
\end{minipage} 
\caption{\label{muF} Modulation factors as function of energy for the four polarimeters.
For Design 1 the modulation factors are shown for two trigger conditions.
For Design 2 the modulation factors are shown when summing the
$\phi$-distributions over all event types (solid line) and for individual
event types (other line styles, see legend).}
\end{center}
\end{figure}
The modulation factors exhibit only a weak energy dependence (Fig.\ \ref{muF}).\\[2ex]
%
%
\begin{figure}[tb]
\begin{center}
\includegraphics[width=3.5in]{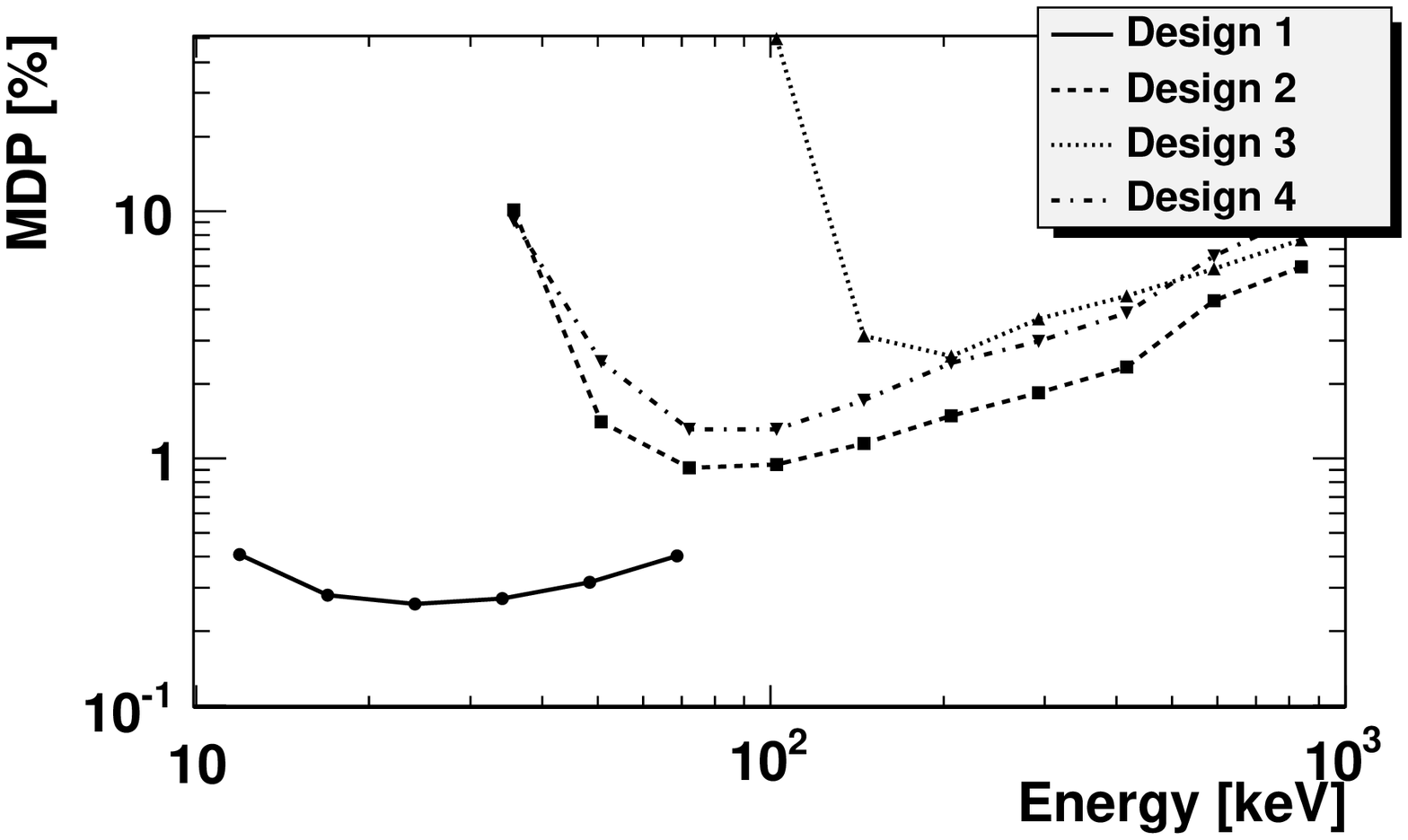}
\caption{\label{mdpF} Differential MDPs of the four polarimeters.
The binning was chosen to have six statistically independent bins 
between 10 keV and 80 keV (Design 1) and ten statistically independent 
bins between 30 keV and 1 MeV (Designs 2-4). For all four designs the binning 
corresponds to a $\Delta$log$_{10}\,E$ of 0.15 per energy bin. 
The lines between the data points are only shown to guide the eye.}
\end{center}
\end{figure}
In the following we present the sensitivities of the four polarimeters for a 100 ksec observation
of a source with a Crab-like flux and energy spectrum. The integral MDPs achieved with the four 
detector configurations are compiled in Table \ref{res}. It should be noted that the MDPs are only 
valid for the case that the signal strongly dominates over the background. 
Design 1 achieves the lowest MDP (0.13\%) followed by Design 2 (0.48\%), Design 3 (0.73\%), 
and Design 4 (1.7\%). The energy resolved MDPs are shown in Fig.\ \ref{mdpF}. 
One clearly sees that Design 1 is a low-energy instrument achieving the 
lowest MDP values at 24 keV. Designs 2 and 3 are most sensitive 
at $\sim$70 keV, and Design 4 at 200 keV.

If Design 1 is used with the requirement of a CZT energy deposition exceeding 10 keV and
a scintillator energy deposition exceeding 2 keV the detection rates drops from 
431 Hz to 149 Hz and the MDP increases from 0.13\% to 0.20\%.

%
%
The background is expected to be lowest for the scintillator-CZT detector assembly (Design 1)
in the focal plane of a Wolter-type mirror assembly: the narrow-FoV makes it possible
to shield the detector assembly with a solid angle coverage close to $4\pi$. Furthermore, the 
total mass of the detectors is much smaller than for the other three designs as the photon 
collection area is provided by the mirror assembly and not by the detectors themselves.
The large mass of the detectors of Designs 2-4 unavoidably scatters high-energy 
photons and particles and leads to a comparatively high background. 

We derived order of magnitude estimates of the expected background rate for 
Design 1 and for Designs 2-4. The results can be used to get a rough estimate for which 
detector configurations and source strengths the background may indeed be negligible.  
Our background estimates for Design 1 assume three identical mirror/polarimeter assemblies.
Each mirror assembly has an effective mirror area of 533 cm$^2$, a maximum mirror diameter 
of 40~cm, and a focal length of 10~m. Each of the three mirror assemblies could be
similar to the two mirror assemblies used for the NuSTAR mission \citep{Harr:10}. 
The three mirror assemblies focus the X-rays onto three identical polarimeters of Design 1.
Assuming an aperture of 4$^{\circ}$ diameter (similar to the one used of NuSTAR), we estimate
a very low 10-80 keV CXB aperture flux count rate. 
The count rate from CXB photons being focussed by the mirrors onto the scattering 
slab is somewhat larger but still below 1/1000 of the count rate produced by a 
1 Crab source. The dominant background will probably be internal 
backgrounds and the shield leakage background. Using the 10 keV - 80 keV background estimate of 
3$\times 10^{-4}$ s$^{-1}$ cm$^{-2}$ keV$^{-1}$ from the simulations of the 
NuSTAR experiment \citep{Harr:10}, we estimate that the high-background event sample 
(no scintillator coincidence required) will have a total background count rate 
corresponding to the rate of Compton scattered events from a 17 mCrab source.
This estimate accounts for all the background events in 96 CZT detectors of the three 
identical polarimeters, and assumes a combination of active and passive 
shielding and the reduction of background events based on the reconstruction of 
the depths of the interactions, similar as used for NuSTAR. 
We conclude that the high-background event  sample will be signal dominated for sources 
with fluxes exceeding $\sim$20 mCrab. The low-background event sample (with a coincident 
trigger in the scintillator) will have a much lower background. Simulations are 
underway to determine this background rate.

For Designs 2-4 the CXB aperture background flux will be relatively low -- if the detector 
assemblies are used with a narrow field of view collimator assembly. 
{\it ASTRO-H} will use a collimator with an aperture of $0.55^{\circ}\times 0.55^{\circ}$
at low energies ($<$150 keV) and with an aperture of $10^{\circ}\times 10^{\circ}$
at high energies ($>$150 keV) \citep{Taka:08}. For the smaller aperture, the 50-150 keV 
CXB aperture flux corresponds to a flux of 0.6 mCrab. For the larger aperture, 
the 150 keV-300 keV CXB aperture flux gives the same count rate as a source with 
a flux of 108 mCrab. The dominant background of designs 2-4 will probably come from 
internal backgrounds and shield leakage. A background rate of 3$\times$10$^{-4}$ 
s$^{-1}$ cm$^{-2}$ keV$^{-1}$ will produce the same rate of 50-150 keV (150-300 keV) 
events as a source with a flux of 425 mCrab (3 Crab). 
Our estimate of the background count rate is motivated by the simulation results 
for NuSTAR and by our own simulations of detector configurations similar to Design 4.
Takahashi et al. (2008) \citep{Taka:08} estimate a
two orders of magnitude lower internal background rate per unit area 
for the Si and CdTe detector assembly of the {\it ASTRO-H} soft gamma-ray telescope - 
but the estimate still needs experimental verification.
From the discussion it is clear that if such extremely low background rates 
cannot be achieved in orbit, then only bright galactic sources can outshine 
the background.
If the detector assemblies of Designs 2-4 are used in wide field of view instruments, 
then the CXB aperture flux becomes the dominant background.
Assuming a 1 sr field of view, the 50-150 keV and 150-300 keV aperture fluxes
correspond to source strengths of 6.1 Crab and 4.2 Crab, respectively.    
In this case, only bright GRBs can outshine the background.

%
%
The four detector assemblies differ not only in the accessible energy ranges and the achieved sensitivities.
The solid state experiments (Designs 1, 2, and 4) exhibit a much better spectroscopic performance 
than the scintillator-only experiment (Design 3). For thick CZT detectors full width half maximum energy 
resolutions of 2.5 keV at 59 keV, 2.7 keV at 122 keV and 4 keV at 662 keV have been reported \citep{Li:10}.
The Si strip detectors operated at -30$^{\circ}$ achieve $\sim$1 keV energy resolutions.
The narrow FoV polarimeter would achieve a good effective energy resolution even though the
energy resolution of the scintillator is poor. The reason is that $<$80 keV photons lose only  
a small fraction of their initial energy in Compton interactions (compare Fig.\ \ref{ComptonEl}).
The BGO scintillators used in Design 3 have a much poorer energy resolution than the CZT
detectors used in Designs 1, 2 and 4, i.e.\ about $\sim$50 keV FWHM at energies 
of a few 100 keV \citep{Blos:09}.
When used with a coded mask, the solid state polarimeters (Designs 2 and 4) achieve better 
angular resolutions than the scintillator polarimeter (Design 3) as the best possible angular 
resolution scales linearly with the CZT pixel pitch (0.6 mm and 2.5 mm), the Si strip pitch
(1.4 mm), and the scintillator slab pitch (6.35 mm).
It should be noted that this section compared instruments to each other  
with different degrees of complexity, different sizes, masses, and costs.
\subsection{Calibration Issues}
Proper on-ground and in-orbit calibration of the experiments will be crucial 
for achieving performances which are limited by the statistics of the 
detected Compton scattered events. We will address several issues in the following:
the calibration of the detector response to photons, systematic uncertainties owing 
to background events, and other technical challenges.   

%
%
One topic of major importance is the calibration of the energy thresholds of individual 
detector elements. Depending on the employed analysis method, the energy threshold may depend 
on the hardware trigger threshold, on the analysis cuts used 
to select valid events, or on both. 
A detailed analysis should account for the fact that the threshold behavior 
of the $j^{\rm th}$ detector element is given by the detection probability 
$p_{\rm j}(E)$ as function of the energy $E$ deposited in the detector element.
Whereas $p_{\rm j}(E)$ is almost a step function for detectors with
excellent energy resolutions (e.g.\ Si and CZT detectors), it is a slowly varying
function of energy for detectors with poorer energy resolutions (e.g.\
CsI, and much more so for plastic scintillators).
As most sources have steeply falling energy spectra $dN/dE\propto E^{-\Gamma}$ with 
$\Gamma\approx 2-4$, an uncertainty of the energy threshold by a factor of 
$\Delta E/E$ will result in a count rate uncertainty by $\approx$ $(\Gamma-1) \Delta E/E$. 
If one is interested in measurements of the polarization fraction and the polarization
direction in multiple energy bins, then one needs to know the probability 
$q_{\rm j,k}(E)$ that an event depositing the energy $E$ in the $j^{\rm th}$ 
detector element will be recorded and will be reconstructed in the $k^{\rm th}$ 
energy bin. Inferring the functions $q_{\rm j,k}(E)$ will usually require a 
good calibration of the energy scale (absolute offsets and linearity) 
and the energy resolution. The functions $p_{\rm j}(E)$ and $q_{\rm j,k}(E)$ 
can be determined from detailed calibration measurements with a series of 
radioactive sources of known activity. 

For all the designs, the energy threshold of the Compton scatterer is 
of special interest as it determines the lower energy threshold for detecting 
Compton events. For Design 1 one can use a single radioactive 
source with a strong emission line at an energy $E_{\gamma}\approx$ 100~keV to make 
a precision measurement of $p_{\rm j}(E)$ of the scintillator rod  at 
energies $E \ll E_{\gamma}$ by Compton scattering photons off the scintillator 
rod and by detecting the scattered photons with CZT detectors positioned at  
different scattering angles. The Compton formula Equ.\ (\ref{cf}) can be used to 
infer $E$ (or a range of possible $E$-values) when the scattering angle and 
the energy deposited in the CZT detector are known. The comparison of the expected 
and observed rates of Compton events can be used to infer $p_{\rm j}(E)$. 
In practice, the method is a bit more complicated and less unambiguous owing to 
the fact that the suitable radioactive sources (i.e. $^{57}$Co) have 
multiple emission lines. A similar procedure can be used for measuring 
$p_{\rm j}(E)$ for the low-Z plastic scintillators of Designs 3.

%
%
The science payload has to include suitable calibration sources 
to make sure that the ground calibration remains valid during the 
flight of the experiments. One needs to be careful to avoid aging 
scintillators (oxidation of hygroscopic materials) and variations 
of the detector gains owing to temperature and/or magnetic field 
variations during the flight. Although radiation damage is a concern, 
the effect seems to be rather small for recent missions 
(e.g.\ for the Swift BAT and for Fermi).  

%
%
In addition to characterizing the energy response of the detectors,
the active volume of each detector element $V_{\rm j}$ has to be 
measured to allow for an appropriate flat-fielding of the experiment.
In CZT detectors for example, $V_{\rm j}$ is usually smaller for pixels
close to the edges of a detector owing to the electric field line 
geometry inside the detector. In addition, impurities like Te-precipitates
can affect $V_{\rm j}$. Some care has to be taken to disentangle rate 
variations stemming from variable energy thresholds (resulting in
different numbers of detected Compton continuum events) from 
rate variations stemming from partially inactive detector volumes. 
When integrating the detectors into the experiment, the detectors
should be distributed as to minimize systematic biases. 
 
%
%
The response of the entire instrument to X-rays and gamma-rays should be tested 
with polarized and unpolarized sources located at different locations relative 
to the experiment.

%
%
The backgrounds on high-altitude balloon flights and on satellite orbits is 
the next concern. As for all X-ray experiments a proper background subtraction
is crucial for the proper determination of fluxes and energy spectra.
It is highly desirable to have experiments which acquire simultaneous 
ON (source) and OFF (background) data. Only the simultaneous acquisition of 
ON and OFF source data enable strictly differential measurements and 
guarantee that time dependent backgrounds impact both data sets 
at the same time. Experiments which take ON data only, are 
notoriously difficult to calibrate in-orbit. If Designs 2-3 are used in coded 
aperture instruments, ON and OFF data are taken at the same time, allowing 
for proper background subtraction. A possibility to enable simultaneous
ON and OFF measurements for Design 1 would be the addition of a second 
scintillator with a long decay time surrounding the scattering rod. 
Pulse shape discrimination could be used to distinguish between
hits in the central scattering rod and hits in the surrounding
scintillator. An alternative could be the addition of a pixelated 
or cross-strip Si detector surrounding the scattering rod 
close to the focal plane of the mirror assembly.
The Si detector could be used to flag events from the OFF region
and to obtain imaging spectropolarimetric information of extended sources.   

Another concern are spatial gradients of the background intensity 
(or the detector response) which might lead to background counts
producing an azimuthal modulation of the count rate.
Detailed Monte Carlo simulation of the entire satellite need to 
be carried through to estimate the magnitude of such effects and
to modify the design in the case that the effect is strong.
Data taken in-orbit can be used to search for such effects using
observations of presumably unpolarized sources (symmetric galaxy clusters),
empty regions in the sky, or repeated observations of one and the same
sources at different parts of an orbit, or with different 
background conditions.
 
Compton telescopes are particularly sensitive to cross talk between detector
elements producing artificial coincidences. Optical cross talk and 
MAPMT cross talk is an issue for Design 3; charge sharing, weighting
potential, and electronic cross-talk are issues of Design 4 where hits
in adjacent pixels can be mis-classified as Compton events. 
These effects can be studied by testing the polarimeter 
over the entire relevant energy range in the laboratory. 

The discussion in this section shows that the calibration of an X-ray polarimeter
is a major issue. The calibration of experiments with good or excellent
energy resolution and with physically separated low-Z scatterers and high-Z absorbers
(Designs 1 and 2) is probably more straight forward than the calibration of polarimeters
with very poor energy resolution and/or where multiple interactions are detected in
one and the same detector (Designs 3 and 4).
\section{Summary and Discussion}
\label{discussion}
Hard X-ray observations have to cope with the steep energy spectra of most astrophysical sources, 
and thus with lower photon fluxes than those in the soft X-ray band. However, hard X-ray 
polarimetry is an exciting upcoming field owing to several facts:
\begin{itemize}
\item Several phenomena can {\it only} be observed at hard X-rays. Examples are 
the measurement of the high-energy end of the thermal emission from BBHs coming 
from the immediate surrounding of the black holes, observations of polarized cyclotron 
lines of magnetars, and the measurement of the polarization near the high-energy 
cutoffs of magnetars owing to the effect of photon splitting.
\item Almost all science topics that can be addressed with soft X-ray polarimetry 
benefit greatly from {\it simultaneous soft and hard X-ray spectropolarimetric observations}. 
The broadband energy dependence of the polarization degree and direction is crucial 
to verify that the models used to explain the soft X-ray polarization results are 
actually correct. In several cases combined soft and hard X-ray polarimetry observations 
are required to determine the model parameters that affect the interpretation of 
the results obtained in the two bands. A prominent example is the study of
BBH systems: the combined soft and hard X-ray observations are needed 
to constrain the parameters describing the black hole, the accretion disk, 
and the corona.   
\item In some sources the polarization degrees at higher energies are expected to be higher 
than at lower energies owing to the more compact emission regions of hard X-rays.
For some sources the effect may make it easier to measure the hard X-ray polarization than 
the soft X-ray polarization.
\item Hard X-rays allow us to study heavily obscured sources with column densities exceeding
10$^{24}$ cm$^{-2}$.  
\item For very hard sources (i.e. hard GRBs) hard X-ray observations achieve similar
MDPs as soft X-ray observations as the photon number only depends logarithmically on the
low energy threshold.
\end{itemize}        
\begin{table}[t]
\begin{tabular}{p{5.2cm}|p{1.6cm}p{1.6cm}p{1.6cm}p{1.6cm}}
\hline
					& Design 1 & Design 2 & Design 3 & Design 4\\ \hline \hline
BBH thermal disk emission		& \hspace*{0.6cm}\checkmark & \hspace*{0.6cm}- & \hspace*{0.6cm}- & \hspace*{0.6cm}- \\
BBH coronal emission			& \hspace*{0.6cm}\checkmark & \hspace*{0.6cm}\checkmark & \hspace*{0.6cm}\checkmark & \hspace*{0.6cm}\checkmark\\
X-ray/$\gamma$-ray pulsars	& \hspace*{0.6cm}- & \hspace*{0.6cm}\checkmark & \hspace*{0.6cm}\checkmark & \hspace*{0.6cm}\checkmark\\
NS cyclotron lines		        & \hspace*{0.6cm}\checkmark & \hspace*{0.6cm}- & \hspace*{0.6cm}- & \hspace*{0.6cm}- \\
NS vacuum birefringence	        	& \hspace*{0.6cm}\checkmark & \hspace*{0.6cm}- & \hspace*{0.6cm}- & \hspace*{0.6cm}- \\
Magnetar X-ray tails		& \hspace*{0.6cm}\checkmark & \hspace*{0.6cm}\checkmark & \hspace*{0.6cm}\checkmark & \hspace*{0.6cm}
\checkmark\\
Magnetar photon splitting		& \hspace*{0.6cm}- & \hspace*{0.6cm}\checkmark & \hspace*{0.6cm}\checkmark & \hspace*{0.6cm}\checkmark\\
AGN coronae 				& \hspace*{0.6cm}$\sim$ & \hspace*{0.6cm}- &\hspace*{0.6cm}-&\hspace*{0.6cm}-\\
Blazar jets					& \hspace*{0.6cm}\checkmark & \hspace*{0.6cm}\checkmark & \hspace*{0.6cm}\checkmark & \hspace*{0.6cm}\checkmark\\
GRB jets					& \hspace*{0.6cm}$\sim$$^a$ & \hspace*{0.6cm}\checkmark & \hspace*{0.6cm}\checkmark & \hspace*{0.6cm}\checkmark\\
Solar Flares				& \hspace*{0.6cm}$\sim$$^b$ & \hspace*{0.6cm}\checkmark & \hspace*{0.6cm}\checkmark & \hspace*{0.6cm}\checkmark\\
Stellar Flares				& \hspace*{0.6cm}- & \hspace*{0.6cm}- & \hspace*{0.6cm}- & \hspace*{0.6cm}-\\
LIV							& \hspace*{0.6cm}\checkmark & \hspace*{0.6cm}\checkmark & \hspace*{0.6cm}\checkmark & \hspace*{0.6cm}\checkmark\\
\hline
\end{tabular}
\hspace*{1cm}$^a$ Requires alert by another instrument.\\
\hspace*{1cm}$^b$ A narrow field of view instrument would be unlikely to detect
exceptionally strong solar flares without alerts from other instruments.\\
\caption{\label{sum} Science topics that can be addresses with the four different instrument
designs.}
\end{table}
Table \ref{sum} shows the science topics that can be addresses with the different instrument
designs. The table accounts for the different energy ranges of the four polarimeters and for their 
different field of views (Design 1: narrow field of view, Designs 2-4: possibly wide field of views). 
An estimate of the actual numbers of detected sources requires careful optimization of the shielding 
concept and is outside the scope of this paper.  

In the light of the results shown in the table three {\it GEMS} and {\it ASTRO-H} follow-up missions
are attractive. The first mission is a narrow FoV broadband X-ray polarimetry mission 
with good sensitivity in the 0.1 keV-2 keV energy regime, an order of magnitude improved 
sensitivity and/or imaging spectropolarimetric capabilities in the 2 keV-10 keV energy band, 
and with spectropolarimetric coverage up to 80 keV. The second mission is a wide FoV 
observatory for spectropolarimetric studies of GRBs and flaring galactic sources, e.g.\  
similar to EXIST, a large HX-POL, or POET. There is a niche for a large area detector 
assembly (similar to HX-POL, GRAPE) used with a pencil beam collimator to measure 
the $>$80 keV polarization properties of galactic sources, e.g.\ BBHs and the various flavors of neutron stars.

The Washington University hard X-ray group is presently assembling a polarimeter 
called {\it X-Calibur} which adopts Design 1. We will report on detailed 
calibration measurements and on comparisons of simulated and experimentally 
measured data in a forthcoming paper. Pending approval by NASA, the polarimeter will 
be flown on a one-day balloon flight in the focal plane of an 
{\it InFOC$\mu$S} mirror assembly \citep{Ogas:05} with $\sim$40cm$^2$ detection area 
in spring 2012 and on subsequent longer balloon flights from Australia. 
In a collaboration with the Naval Research Laboratory (E.\ Wulff et al.), 
the group is also testing prototypes of the {\it HX-POL} Si-CZT polarimeter (Design 2). 
\\[2ex]
{\large \bf Acknowledgements:} 
The authors thank Martin Israel, Jonathan Katz (both Washington Univ.\ in St.\ Louis), 
Paolo Coppi (Yale), and J. Schnittman (Goddard Space Flight Center) for reading 
the manuscript carefully and for very valuable suggestions. Insightful comments by an 
anonymous referee helped to strengthen the paper substantially. 
HK acknowledges NASA for support from the APRA program under the grant NNX10AJ56G 
and support from the high-energy physics division of the DOE. The Washington University
group is grateful for discretionary funding of the {\it X-Calibur} polarimeter 
by the McDonnell Center for the Space Sciences. 

\end{document}